\documentclass[conference,letterpaper]{IEEEtran}
\usepackage[T1]{fontenc}
\usepackage{amsmath,amssymb}
\usepackage{booktabs,array,graphicx}
\usepackage{cite}
\usepackage{microtype}
\usepackage{etoolbox}
\usepackage[hidelinks]{hyperref}
\patchcmd{\abstract}{---\relax}{:\ \relax}{}{\PackageError{manuscript}{Abstract label patch failed}{}}
\patchcmd{\IEEEkeywords}{---\relax}{:\ \relax}{}{\PackageError{manuscript}{Keywords label patch failed}{}}

\hypersetup{pdftitle={Retained Encrypted Search with Certified Split Joins},
  pdfauthor={Yavor Litchev, Liwen Ouyang},
  pdfsubject={Retained exact search, certified split joins, and bounded adaptive Ring-LWE security}}
\newtheorem{theorem}{Theorem}[section]
\newtheorem{definition}[theorem]{Definition}
\newtheorem{lemma}[theorem]{Lemma}
\newtheorem{proposition}[theorem]{Proposition}
\newtheorem{corollary}[theorem]{Corollary}
\newenvironment{proof}{\par\noindent\emph{Proof.}}{\hspace*{\fill}\nobreakspace$\square$\par}
\newcounter{protocol}
\newenvironment{protocol}[1]{%
  \begin{figure}[htbp]\refstepcounter{protocol}\hrule\medskip
  \noindent\textbf{Algorithm \theprotocol.\ #1}\par\smallskip\small\raggedright
}{\medskip\hrule\end{figure}}
\newcommand{\Ham}{\operatorname{Ham}}
\newcommand{\Top}{\operatorname{Top}_3}
\newcommand{\TopSum}{\operatorname{TopSum}}
\newcommand{\Finish}{\mathsf{Finish}}

\newcommand{\Expand}{\mathsf{Expand}}
\newcommand{\Prep}{\mathsf{Prepare}}
\newcommand{\Hash}{\mathsf{Hash}}
\newcommand{\Adv}{\operatorname{Adv}}
\newcommand{\CBD}{\operatorname{CBD}}

\newcommand{\ceil}[1]{\left\lceil#1\right\rceil}
\newcommand{\norm}[1]{\left\lVert#1\right\rVert}
\newcommand{\Base}{\mathsf{K}}
\newcommand{\Head}{\mathsf{H}}
\newcommand{\Generation}{\mathsf{G}}
\newcommand{\State}{\mathcal{S}}
\newcommand{\History}{\mathcal{U}}
\newcommand{\Ledger}{\mathcal{J}}
\newcommand{\Ready}{\mathcal{R}}
\newcommand{\Pending}{\mathcal{P}}
\newcommand{\Spent}{\mathcal{C}}

\newcommand{\CombQPaidSeconds}{12.159}
\newcommand{\CombQFirstMs}{243.878}
\newcommand{\CombQPrepareSeconds}{3.514}
\newcommand{\CombQPublicMs}{217.625}
\newcommand{\CombQPrivateMs}{25.079}
\newcommand{\CombQSetupMiB}{300.75}
\newcommand{\CombQRetainedMiB}{3.0}
\newcommand{\CombQGalleryMiB}{512}

\newcommand{\CombDetPaidSeconds}{14.228}
\newcommand{\CombDetFirstMs}{239.789}
\newcommand{\CombDetPrepareSeconds}{7.205}
\newcommand{\CombDetPublicMs}{213.312}
\newcommand{\CombDetPrivateMs}{25.145}
\newcommand{\CombDetSetupMiB}{186.56}

\newcommand{\CombStatPaidSeconds}{9.825}
\newcommand{\CombStatFirstMs}{156.139}
\newcommand{\CombStatPrepareSeconds}{4.635}
\newcommand{\CombStatPublicMs}{116.155}
\newcommand{\CombStatPrivateMs}{39.635}
\newcommand{\CombStatSetupMiB}{186.06}
\newcommand{\CombStatRetainedMiB}{1.5}
\newcommand{\CombStatGalleryMiB}{256}
\newcommand{\CombStatBodyKiB}{192}

\newcommand{\CombQPaidRatio}{1.237}
\newcommand{\CombQPaidRatioMin}{1.229}
\newcommand{\CombQPaidRatioMax}{1.262}
\newcommand{\CombQFirstRatio}{1.569}

\newcommand{\CombDetPaidRatio}{1.443}
\newcommand{\CombDetPaidRatioMin}{1.438}
\newcommand{\CombDetPaidRatioMax}{1.464}
\newcommand{\CombDetFirstRatio}{1.536}

\newcommand{\CombPaidProfileTable}{
\begin{tabular}{lrrrr}
\toprule
Profile & Prep. (s) & Public (ms) & Private (ms) & Paid (s) \\
\midrule
Q120 fixed24 & \CombQPrepareSeconds & \CombQPublicMs & \CombQPrivateMs & \CombQPaidSeconds \\
P64 fixed24 & \CombDetPrepareSeconds & \CombDetPublicMs & \CombDetPrivateMs & \CombDetPaidSeconds \\
P60 residual & \CombStatPrepareSeconds & \CombStatPublicMs & \CombStatPrivateMs & \CombStatPaidSeconds \\
\bottomrule
\end{tabular}
}

\title{Retained Encrypted Search with Certified Split Joins}
\author{
  \IEEEauthorblockN{Yavor Litchev}
  \IEEEauthorblockA{\href{mailto:ylitchev@stanford.edu}{ylitchev@stanford.edu}}
  \and
  \IEEEauthorblockN{Liwen Ouyang}
  \IEEEauthorblockA{\href{mailto:liwen@xtrace.ai}{liwen@xtrace.ai}}
}

\begin{document}
\maketitle

\begin{abstract}
Retaining encrypted search across data edits and worker replacement requires
separating computational validity from current release authority. We give exact
Hamming search with at most one authorized decryption and decoding attempt per
query preparation. Under authenticated prescribed execution, exclusive
nonrollback owner state, an ideal public seed-expansion interface and a stated
observation model, we reduce bounded adaptive privacy to ordinary ring learning
with errors (Ring-LWE). Expanded second components depend only on public masks
and fixed maps, independently of queries and the fresh database key. This
compiler invariant permits a retained split join against a smaller-modulus
encrypted database. A simultaneous bound on the completed circuit covers a
finite honest reservation family, including adaptive queries and slot selection,
with explicit statistical correctness and privacy losses. In three
post-calibration local CPU comparisons on 32,768 vectors of 512 bits, the
60-bit-join construction takes a median \CombStatPaidSeconds{} s for an
end-to-end two-query workload, versus \CombQPaidSeconds{} s for the optimized
full-source-modulus baseline. These times include setup, transfers, preparation
and private completion. They are local implementation results, not deployment
or concrete-security claims.
\end{abstract}

\section{Introduction}
\label{sec:introduction}
Encrypted exact search hides database rows and queries from the storage host
while outsourcing Hamming-distance evaluation. Keys and the encrypted database
serve multiple queries, but each query-independent preparation permits at most
one issuance. The owner retains a private mask and noise, while the server
retains matching prepared ciphertexts. \emph{Private completion} is the owner's decryption
and decoding of an authenticated complete result.

We study retained search under an active host and an integrity-protected
executor holding encrypted inputs and confined signing authority, but no
homomorphic-encryption (HE) secrets. It authenticates complete prescribed
results before private completion, addressing the reaction-attack risk from
malicious ciphertexts~\cite{vfhe}. The owner retains truthful history and may
keep all plaintext data, making plaintext caching a permitted control.

The owner corrects scores from immutable computation base $\Base$ to the current
data head $\Head$ at serialized release. Reconstruction changes worker generation
$\Generation$ and transfers only held unused preparations. Durable records
prevent repeated issuance or completion. We instantiate this protocol using
projected Brakerski--Gentry--Vaikuntanathan (BGV) queries.

Can a retained preparation be combined with a reference database encrypted at
a smaller modulus while preserving these guarantees? We call this encrypted
reference database the
\emph{gallery}, and the sum of query--column ciphertext products the \emph{join}.
Separate rounding of the retained and late first components is not additive,
so we bound the \emph{completed} split circuit rather than its incomplete
preparation. The expansion compiler translates coefficient-extraction maps
into key switching, with second components independent of the query and fresh
gallery secret. This invariant permits a simultaneous rounding bound over all
reserved seeds, covering adaptive choices.

We make two contributions.
\begin{enumerate}
 \item A retention-preserving exact-search protocol and security composition.
 Its current-output invariant couples unique projected issuance, authenticated
 whole-result admission, durable consumption, correction and reconstruction.
 An acyclic independent-recipient reduction handles delayed adaptive queries
 under ordinary Ring-LWE and explicit lifetime accounting.
 \item A certified retained split join based on the second-component compiler
 invariant and a finite-family rounding bound. Its exact output contract and
 statistical loss compose with the same owner authority. The implementation
 measures full-lifetime local costs.
\end{enumerate}

Offline/online HE computation and one-query seeds have recent examples in
ZipPIR and FlashPIR~\cite{zippir,flashpir}. Attested or verified evaluation appears in
Argos and vFHE~\cite{argos,vfhe2024}. Current authenticated data also underlies
dynamic verification~\cite{benabbas,integridb}. BGV modulus switching, hoisting,
Bernstein concentration and finite union bounds are established tools.
Our contribution is their composition with a retained split join, adaptive
selection and one-use current release, not a new probability inequality or
optimality result. Evaluation compares complete construction profiles rather than an
isolated optimization.

\section{Functionality and Execution Model}
\label{sec:model}
\subsection{Exact Current Search}
Let $D_0\in\{0,1\}^{n\times d}$ be the initial database of $n\ge3$ binary
vectors of dimension $d\ge1$, with ordered distinct row identifiers
$\mathcal I=(\mathrm{id}_0,\ldots,\mathrm{id}_{n-1})$.
A query is $q\in\{0,1\}^d$. A truthful authorized edit
$u_\ell=(i_\ell,j_\ell,b_\ell,b'_\ell)$ satisfies
$D_{\ell-1}[i_\ell,j_\ell]=b_\ell$ and $b'_\ell=1-b_\ell$.
Its application produces $D_\ell$. Repeated coordinates and reversions are
allowed. Let $\History_e=(u_1,\ldots,u_e)$ be the ordered edit log and
$\Head_e=(e,h_e)$ its endpoint identifier. For a fixed base and a single
nonrollback history, the counter $e$ identifies the prefix. The optional
metadata $h_e$ is a fresh opaque owner label, or a fixed public placeholder
if unused. Reversions advance the head even when they restore earlier contents.

\begin{definition}[Current-search functionality]
At a serialized release with head $\Head_e$, define
\begin{align}
 v_e(q)[i]&=\Ham(D_e[i],q)
          =\sum_{j=0}^{d-1}(D_e[i,j]\mathbin{\oplus}q[j]),\label{eq:ham}\\
 \mathcal F(D_e,q,\mathcal I)&=(\Head_e,v_e(q),\Top(v_e(q),\mathcal I)).
 \label{eq:functionality}
\end{align}
$\Top$ sorts rows lexicographically by $(v_e(q)[i],i)$ and returns the first
three distances and their bound identifiers. The owner may release just the
top three.
\end{definition}
Answers are current at release. Later edits do not invalidate them. Ranking
is private. Associated-record retrieval and encrypted server-side top-$k$
are separate functionalities.

Encrypted evaluation uses the sign map $\chi(b)=1-2b$. For
$x,q\in\{0,1\}^d$, the sign-correlation identity is
\begin{equation}
 \Ham(x,q)=\frac{d-\sum_{j=0}^{d-1}\chi(x_j)\chi(q_j)}{2}.
 \label{eq:sign}
\end{equation}

\subsection{Parties and Trust Boundary}
In a client-server deployment, source and owner roles may share one client,
while the server runs the worker.
The honest source encrypts $D_0$, authenticates its ordered identifiers and
layout, and provisions the owner's private roles. The owner holds HE secrets,
queries, and truthful edit history, and may retain all plaintext data and
prior scores. One serialized owner authority governs issuance and release,
with faithful storage of monotone reservation and consumption records.

The worker's executor $\mathcal E$ is an integrity-protected interface for
computation and signing. It holds the public key graph, encrypted
index, approved evaluation, a \emph{codec} (encoding and decoding rule), and
a confined receipt-signing key.
On authenticated requests, it runs only the bound computation and signs only
complete prescribed results, with no arbitrary-result signing interface.
The untrusted host controls external storage, packet delivery, scheduling,
replay, omission, and public worker-state cloning.

Enrollment is the authenticated registration that binds $\mathcal E$'s
verification key and generation to the
prescribed program and assets. Attested execution can instantiate this
interface, or sound verification can replace it with its corresponding error
term. An unconstrained server
signature does not suffice.
The source and owner are honest, the executor's signing key remains confined,
and the owner authority is exclusive and nonrollback. Private provisioning and
plaintext transport require faithful confidential authenticated boundaries.
Availability and recovery of lost private RAM state are not required.

\subsection{Observations and Adaptive Inputs}
\begin{definition}[Public leakage $\mathcal L$]
The host observes dimensions and packing layout, ordered identifiers, output
ciphertext type and wire profile (response modulus, coefficient width and codec),
public encrypted assets, ciphertext-derived bindings, message lengths,
head counters, optional opaque head labels, generations, reservation identities,
routing, and the logical
schedule of preparation, issuance, updates, rejection, finalization,
reconstruction, and cache selection. Public policy outcomes are included.
\end{definition}
Rows, query words, edit contents, scores, and private masks remain in the
owner/source boundary. Cache hits or avoided traffic may disclose equality
and reuse patterns. These observations are included in the leakage. Secret-dependent
execution durations, memory addresses, residual private buffers, and
microarchitectural observations are outside the observation model. Host-chosen
delivery delays remain observable. The computational theorem therefore
concerns logical protocol events. It does not claim that the current
variable-time implementation hides private execution time.

An honest input policy is the client's operation-selection procedure. Its
interface exposes data and earlier authorized answers, but not HE secrets,
hidden phases or private encryption randomness. We compare paired policies that are \emph{causally
$\mathcal L$-compatible}: given any common public transcript, they propose
valid private operations with the same next public event, dimensions, layout, routing,
lengths, and public decision. Private ideal answers may differ. An incompatible
proposal terminates both worlds with the same neutral observation.
Ciphertext hashes are computed from each game's actual bytes. Compatibility
does not demand equality of hashes from different games.

\section{Encrypted Representation and Retained Evaluation}
\label{sec:arithmetic}
The baseline expands queries and joins database columns at $Q$. Response
compression uses $p$ (Section~\ref{sec:compression}). The split join adds an
intermediate modulus $P$ (Section~\ref{sec:retainedjoin}), giving
$Q\to P\to p$ at unchanged ring dimension.
\subsection{An Independent Recipient Graph}
Let $m=\smash{2^{\ceil{\log_2d}}}$ be the padded query length, the smallest
power of two at least $d$. Choose an odd plaintext modulus $t>2m$.
Since $m\ge d$, every sign correlation in $[-d,d]$ has a unique centered
representative modulo $t$, allowing exact recovery from its plaintext residue.

Let $N=2^L\ge2m$ be the ring dimension with $m\mid N$, and $r=N/m$ the
projection stride. For any integer modulus $u>1$, define
\[
 R=\mathbb Z[X]/(X^N+1),\qquad R_u=R/uR.
\]
The source ciphertext modulus $Q>1$ is odd, with
$\gcd(t,Q)=\gcd(t,N)=1$.
We use coefficient $\ell_1$, Euclidean and maximum norms, denoted
$\norm{\cdot}_1$, $\norm{\cdot}_2$ and $\norm{\cdot}_\infty$. Products are
negacyclic, with $X^N=-1$. We use the leveled BGV scheme~\cite{bgv}.
Each secret has independent uniform coefficients in $\{-1,0,1\}$.
For an integer $\eta\ge1$, each error coefficient has centered binomial
distribution (CBD)
$\CBD_\eta=\sum_{j=1}^{\eta}(U_j-V_j)$ for independent fair bits.
We call a fresh ciphertext a \emph{parent}, and its decryption key its
\emph{recipient}. A recipient is a key role, not an additional protocol
participant. A fresh secret-key ciphertext is sampled as
\begin{equation}
 C=(c_0,c_1)=(\mu+te-as,\ a)\in R_Q^2.
 \label{eq:parent}
\end{equation}
Here $\mu$ has centered plaintext coefficients modulo $t$, $e$ is the error
polynomial, $s$ is secret, and $a$ is an independent uniform public mask in $R_Q$.
The phase $\Phi_s(C)=c_0+c_1s\equiv\mu+te\pmod Q$
cancels the mask. An integer lift represents this residue with integer
coefficients, not necessarily centered, and must be bounded before decoding.

Sample independent secrets $s_0,\ldots,s_L$ and, when needed, a final $s_f$,
with $f=L+1$. In this sequential graph, $g=L$ denotes the gallery key role.
Public key-switching keys, called \emph{hints}, contain encrypted
secret-derived terms used to change a ciphertext's recipient without decryption.
At each level $\ell<L$, publish identity and automorphism hints
addressed to $s_{\ell+1}$. For an integer gadget radix $\beta\ge2$ and
$k=\ceil{\log_\beta Q}$, a column with target $\tau$ is
\begin{equation}
 (b_j,a_j)=(\beta^j\tau+te_j-a_js_{\ell+1},a_j),
 \quad 0\le j<k,
 \label{eq:hint}
\end{equation}
where $a_j$ and $e_j$ are fresh mask and error draws, $\tau=s_\ell$ or
$\tau=\sigma_\ell(s_\ell)$ and
$\sigma_\ell(f)(X)=f(X^{1+N/2^\ell})$.
Final hints use the same sampling rule with recipient $s_f$ and target
$s_g$ or $s_g^2$, as required by the terminal type below.
Each role also publishes one independent parent with $\mu=0$ as its public-key
record. It counts as an exposed sample, although queries and gallery entries
use secret-key encryption.

For $w=\sum_{j<k}\beta^jd_j\bmod Q$, use coefficientwise base-$\beta$ digits
of representatives in $[0,Q)$. With hints addressed to $s_v$, define
\begin{equation}
 \mathsf{KS}_{\tau\to v}(w)=
 \left(\sum_jd_jb_j,\ \sum_jd_ja_j\right).
 \label{eq:ks}
\end{equation}
Its phase under $s_v$ is $w\tau+t\sum_jd_je_j$ modulo $Q$.
This graph uses independent leveled switching
as in BGV~\cite{bgv}. Its key-dependency graph is acyclic: every hint target
depends only on keys earlier than its recipient.

\subsection{Projected Queries}
We encode and mask a compact query, expand it into encrypted signs, and compute
database correlations.
For $f=\sum_i f_iX^i$, projection
$\pi_r(f)=(f_0,f_r,\ldots,f_{(m-1)r})$ extracts $m$ coefficients at stride $r$.
The embedding $\iota_r$ reverses this placement, inserting zeros elsewhere.
Let $\alpha\in\{1,\ldots,t-1\}$ be the canonical inverse of $N$ modulo $t$,
which compensates for the expansion factor $N$, and encode the query as
\begin{equation}
 \mu_q=\iota_r\bigl(\alpha\chi(q_0),\ldots,\alpha\chi(q_{d-1}),
                   0,\ldots,0\bigr),
 \label{eq:querymessage}
\end{equation}
with coefficients centered modulo $t$. A fresh public seed $z$ yields uniform
$a_z\in R_Q$. The owner prepares the private mask $M_z$ and issues the masked first-component row:
\begin{equation}
 M_z=\pi_r(a_zs_0),\qquad
 x_z(q)=\pi_r(\mu_q)+t e_z-M_z\bmod Q.
 \label{eq:query}
\end{equation}
Only $(z,x_z(q))$ is sent. The secret $s_0$ still has $N$ coefficients.
The transmitted row has the same distribution as the selected first-component
coefficients of a fresh ciphertext in the full $N$-dimensional ring.
The private noise row $e_z\in\mathbb Z^m$
has independent $\CBD_\eta$ coefficients. The owner holds the query-independent
$(z,M_z,e_z)$ preparation until a single issuance.

For $x=(x_0,\ldots,x_{m-1})\in(\mathbb Z/Q\mathbb Z)^m$, define
$\Expand(x,a_z)$ from $(\iota_r(x),a_z)$ using $\log_2r$ sum-only
algebraic trace levels, then $L-\log_2r$ two-way expansion levels. Level
$\ell$ switches identity and automorphic branches to the same new recipient,
then adds them. Two-way levels also
subtract them, shifting the difference by $X^{-2^\ell}$. Lower outputs precede upper outputs.
Its $m$ output ciphertext pairs, called \emph{leaves}, under $s_g$
encode query signs as constants for
$x=x_z(q)$.

\begin{lemma}[Projection and retained affine identity]
\label{lem:affine}
In error-free phase algebra, leaf $j$ is $N$ times the selected input phase
coefficient at $rj$. Unselected coefficients cancel. With fixed hints and
$a_z$, the complete component identity is
\begin{equation}
 \Expand(x,a_z)_j=\Expand(0,a_z)_j+(Nx_j,0).
 \label{eq:affine}
\end{equation}
\end{lemma}
\begin{proof}
On support at multiples of $2^\ell$, $\sigma_\ell$ multiplies coefficient
$i$ by $(-1)^{i/2^\ell}$. The sum selects and doubles even positions, while
the shifted difference selects and doubles odd positions. Induction
eliminates coordinates outside multiples of $r$, then isolates $rj$ in source
order, with total multiplier $2^L=N$. Switching decomposes only the second
component and adds the transformed first component afterward. Its second
component is therefore independent of the original first component, whose
subsequent additions and shifts are linear. The same induction on
$\iota_r(x)$ proves Equation~\eqref{eq:affine} exactly in $R_Q$.
\end{proof}
Here $\Expand(0,a_z)$ has a zero input first component, rather than zero
plaintext. It depends only on the seed and fixed switching hints.

Partition the database into $g_n=\ceil{n/N}$ groups of at most $N$ rows.
Within each group, encrypt feature column $A_j(X)=\sum_i\chi(D_0[i,j])X^i$
under $s_g$, with local row indices $0\le i<N$ and zero padding.
We suppress the group index and use $D_0[i,j]$ for its local rows.
Decoded groups are concatenated in original row order.
Write its encrypted pair as $B_j=(B_{j0},B_{j1})$. Multiplying each column
by its query-sign leaf and summing places row $i$'s sign correlation at
coefficient $i$, for conversion to distance by Equation~\eqref{eq:sign}.
Use $(u,v)\otimes(b,c)=(ub,uc+vb,vc)$ for unrelinearized pair multiplication.
Let $T_z=(T_0,T_1,T_2)=\sum_{j<d}\Expand(0,a_z)_j\otimes B_j$ be the
server's query-independent prepared \emph{tensor}, a three-component product ciphertext.
The arriving first-component coefficient row supplies
the late correction
\begin{equation}
 \Delta T=\left(\sum_{j<d}Nx_jB_{j0},\
                   \sum_{j<d}Nx_jB_{j1},\ 0\right).
 \label{eq:late}
\end{equation}
The complete tensor $T_z+\Delta T$ uses decryption multipliers
$(1,s_g,s_g^2)$. Its unchanged quadratic component can be switched in
advance, using input equality rather than additive gadget switching.

\subsection{Typed Terminals and Phase Bounds}
A \emph{terminal} is a completed output ciphertext. The three types encode the
same correlations, trading final switching work against component count and
decoding requirements. Their decryption multipliers are
\begin{equation}
\begin{aligned}
 \boldsymbol w_{\rm pair}&=(1,s_f),&
 \boldsymbol w_{\rm mixed}&=(1,s_g,s_f),\\
 \boldsymbol w_{\rm tensor}&=(1,s_g,s_g^2).
\end{aligned}
 \label{eq:types}
\end{equation}
Pair switches both linear and quadratic tensor terms to $s_f$.
Mixed switches only the quadratic term and retains the linear term under
$s_g$. Tensor retains the unrelinearized product and needs no final role.
Pair performs its linear switch after Equation~\eqref{eq:late}.
Mixed and tensor need no online terminal switch.
An ordinary $(1,s,s^2)$ decoder cannot decode the mixed terminal.
For each type, the phase is $\sum_i C_iw_i$ modulo the ciphertext modulus,
where $C_i$ are the terminal components.

Define the fresh phase bound $F$, single-switch error bound $S$, complete
expanded-leaf bound $E$, and terminal phase bound $B_\nu$ by
\begin{align}
 F&=\lfloor t/2\rfloor+t\eta,&
 S&=tN\eta k(\beta-1),\notag\\
 E&=NF+2(N-1)S,&
 B_\nu&=NdFE+(2-\nu)S,
 \label{eq:bounds}
\end{align}
for pair, mixed, tensor indices $\nu=0,1,2$ respectively.

\begin{lemma}[Complete source-phase bound]
\label{lem:phase}
Each complete authorized terminal has an integer phase lift $v$ such that
$\norm{v}_\infty\le B_\nu$ and
\[
 v_i\equiv\sum_{j<d}\chi(D_0[i,j])\chi(q_j)\pmod t
\]
on actual rows, with zero plaintext on padding. If $2B_\nu<Q$, centered
source decryption recovers those plaintexts exactly.
\end{lemma}
\begin{proof}
Canonical digits satisfy $\norm{d_j}_\infty\le\beta-1$, so each switching
error is bounded by $S$, using
$\norm{uv}_\infty\le N\norm{u}_\infty\norm{v}_\infty$.
Automorphisms and monomial shifts are signed permutations and preserve
the coefficient infinity norm. Each output combines two switching errors.
Their propagated total after $L$ levels is
$2S(1+2+\cdots+2^{L-1})=2(N-1)S$.
Lemma~\ref{lem:affine} eliminates unselected input coordinates before
applying the fresh bound $F$ to the selected ones, giving $E$.
No fresh bound is assigned to the sparse restored phase or the zero-first
preparation alone.

Gallery phases have bound $F$. Their $d$ negacyclic products with complete
leaves have aggregate bound $NdFE$. The typed final switches add zero, one,
or two errors $S$. Every error is divisible by $t$ in its integer lift.
Since $N\alpha=1\bmod t$, the plaintext is the sign correlation.
The bound below $Q/2$ makes this lift the unique centered phase. Only then
is reduction modulo $t$ applied. Centering that residue gives the integer
correlation in Equation~\eqref{eq:sign}, since $t>2d$.
\end{proof}

\section{Protocol and Correctness}
\label{sec:protocol}
\subsection{Computation, Authority, and State}
$\Base$ binds the encrypted base, independent key graph, ordered identifiers,
layout, terminal type, permitted wire profiles, their selection rule and
prescribed program. $\Head_e$ authorizes
current answers. $\Generation$ names the enrolled worker and its receipt/channel
keys, not the physical server. Correct computation under $\Base$ authorizes no
release at an obsolete head or generation.
An \emph{original} $o=(z,x_z(q))$ is an issued query message.

The authority and capability state is
\begin{equation}
 \State=(\Base,\Head,\Generation,\History,\Ledger,\Ready,\Pending,\Spent),
 \label{eq:state}
\end{equation}
Here $\History$ is the truthful edit log, $\Ledger$ the durable lifetime ledger
charging one slot per preparation attempt, $\Ready$ exclusively held unused
preparations, $\Pending$ pending originals, and $\Spent$ consumed or retired
originals. Sans-serif symbols denote identifiers, calligraphic symbols state
collections. Pending entries bind queries to issue-time metadata:
\begin{equation}
 \gamma_o=(\Base,\Generation_o,\Head_o,n_o,z_o,\ell_o,\Hash(o)).
 \label{eq:binding}
\end{equation}
The fresh nonce $n_o$ names the issued instance, with subscripts $o$ denoting
issue-time values. $\Base$ fixes admissible formats and complete payload extent
for each profile.
Preparation, original and result messages have distinct signed domains.
Identifiers are initially distinct ideal names, realized by random labels
(Section~\ref{sec:security}).
Owner closure invalidates $\Ready,\Pending$ and completion permissions,
halting use without refunds. Durable failure closes before dependent work.

Algorithms~\ref{alg:issue}--\ref{alg:admit} implement serialized owner transitions.
$\Prep(\Base,\Generation,z,\ell)$ authenticates the owner, checks $\Base,\Generation$,
regenerates $a_z$ from $z$, and registers per-group seed-only state from $T_z$
(Section~\ref{sec:arithmetic} or~\ref{sec:retainedjoin}), with permitted quadratic
switching. Its authenticated acknowledgment binds these four arguments.
Plaintext queries, $M_z$ and $e_z$ remain private.
$\Finish(y)$ decrypts all groups of the admitted payload with multipliers
\eqref{eq:types} and the enrolled codec. It checks zero padding and returns
base distances $v_0(q)$ via~\eqref{eq:sign}, without correction, ranking or release.

\begin{protocol}{Private preparation and unique issue}
\label{alg:issue}
\textbf{Input:} open owner state $\State$ and enrolled executor $\mathcal E$.
\begin{enumerate}
 \setlength{\itemsep}{2pt}
 \item \textbf{Prepare:} durably charge a fresh slot $\ell$ in $\Ledger$.
 Draw fresh $z,e_z$ and compute $M_z$ by~\eqref{eq:query}.
 Request $\Prep(\Base,\Generation,z,\ell)$ from $\mathcal E$.
 \item On an authenticated acknowledgment matching the preparation binding,
 store $\ell\mapsto(z,M_z,e_z)$ in $\Ready$.
 Failure closes the owner without refund.
 \item \textbf{Issue $(\ell,q)$:} require a ready slot $\ell$, take its tuple
 $\Ready[\ell]$ and delete that entry before validating $q$ and forming $o=(z,x_z(q))$.
 \item Choose $n_o$, form $\gamma_o$, durably mark $\ell$ issued, and store
 $n_o\mapsto(q,\gamma_o)$ in $\Pending$.
 \item Send the owner-signed $(\gamma_o,o)$.
\end{enumerate}
Any partial failure closes the owner without refund.
\end{protocol}

The executor verifies the bound original and matching preparation, retires the
latter, adds~\eqref{eq:late} or~\eqref{eq:joinretained}, then performs required
switches and canonical encoding. It signs $\gamma_o$, the selected wire profile
and the complete payload under
$\Generation$. Subsets cannot authorize $\Finish$.

\begin{protocol}{Authenticated completion and current release}
\label{alg:admit}
\textbf{Input:} state $\State$, nonce $n_o$ identifying original $o$, and response $y$.
All steps run under the owner's release/edit lock, with current head $\Head_e$.
\begin{enumerate}
 \setlength{\itemsep}{2pt}
 \item Reject and stop unless $n_o$ has an issued, unconsumed entry in $\Pending$.
 Read $(q,\gamma_o)$ and check the complete public frame, signature and binding.
 \item If public admission fails, retire $o$ durably, remove entry $n_o$ from $\Pending$,
 and stop without private completion.
 \item Durably mark $o$ consumed in $\Spent$ and remove entry $n_o$ from $\Pending$
 before any private arithmetic.
 \item If $\Head_o\ne\Head$, $\Generation_o\ne\Generation$, or a cache
 answer has superseded $o$, stop without $\Finish$ or release.
 \item Create and spend an internal one-use permission before $v\gets\Finish(y)$.
 Partial calls cannot be retried.
 \item Compute $v_e\gets v+\Delta_e(q)$ using~\eqref{eq:correction}. Release
 $(\Head_e,v_e,\Top(v_e,\mathcal I))$.
 Private failure closes the owner without release or refund.
\end{enumerate}
\end{protocol}

\subsection{Ordered Local Correction}
Keeping the encrypted base at $D_0$, correct row $i$'s distance by
\begin{equation}
 \Delta_e(q)[i]=
 \sum_{\substack{1\le\ell\le e\\i_\ell=i}}
 \bigl((b'_\ell\oplus q[j_\ell])-(b_\ell\oplus q[j_\ell])\bigr).
 \label{eq:correction}
\end{equation}
\begin{lemma}[Telescoping correction]
\label{lem:correction}
For every truthful ordered history,
$v_0(q)+\Delta_e(q)=v_e(q)$.
\end{lemma}
\begin{proof}
Each edit changes one summand of Equation~\eqref{eq:ham} by its term in
Equation~\eqref{eq:correction}. Summing telescopes, including repeated edits
and reversions.
\end{proof}
Short histories reduce the $O(e)$ scan cost per fresh query, but exactness
requires no sparsity assumption.
The owner advances $\Head$ before dependent work and applies the same row
correction to cached exact score vectors. The baseline corrects edits preceding
issuance but retires a query overtaken by a later edit, despite its valid
cryptographic base. Appendix~\ref{app:editcommuting} permits completion across
in-flight edits under additional live correction checks.
For comparison, server correction can add encrypted feature contributions
before one complete terminal conversion. Independent rounding of each
contribution is a different graph and does not follow from ring linearity.

\subsection{Reconstruction and Permitted Caching}
Reconstruction enrolls a rebuilt or replacement worker, possibly on the same
machine, preserving $\Base,\Head,\Ledger,\Spent$ while advancing $\Generation$
and revoking old pending originals. Only held unused preparations transfer.
They remain unready until the new executor acknowledges
$\Prep(\Base,\Generation,z,\ell)$. Imports and public preparation remain necessary.
Private objects are not recreated from seeds or journals, and lost capabilities
are not restored.

A ready authenticated plaintext cache may answer from exact current rows under
the same authority without issuing an HE original. If acquisition overlaps
a pending HE original, a cache win supersedes it. Later finalization performs no
private completion and retains the charge, preventing a second decryption attempt.

\begin{lemma}[Monotone capability invariant]
\label{lem:oneuse}
For every allowed trace, each private preparation issues at most one original,
and each original invokes at most one whole-result private completion.
Edits, reconstruction, rejection, and cache finalization refund neither.
\end{lemma}
\begin{proof}
Algorithm~\ref{alg:issue} charges before creating $\Ready[\ell]$ and removes it
before fallible issuance. Issued entries never return. Reconstruction transfers
only unused entries. Algorithm~\ref{alg:admit} consumes the pending entry durably
before creating and spending its one-use completion permission. Missing entries
prevent repeat admission, even after partial work. Charges remain spent.
Induction proves all three assertions.
\end{proof}

\begin{definition}[Retention and release invariant]
\label{def:retention}
For a protocol trace prefix $\tau$, let $\mathsf{iss}_\tau(\ell)$ count
originals issued from slot $\ell$ and $\mathsf{fin}_\tau(o)$ count whole-result
private completion attempts for original $o$.
The predicate $\mathsf{Inv}_{\rm ret}(\tau)$ requires all of the following:
\begin{enumerate}
 \setlength{\itemsep}{1pt}
 \item $\mathsf{iss}_\tau(\ell)\le1$ and $\mathsf{fin}_\tau(o)\le1$.
 The charged-slot set and consumed or retired originals grow monotonically.
 \item Each completion follows durable consumption and whole-result admission
 under the exact binding $\gamma_o$. Its issue head and generation equal the
 current $\Head$ and $\Generation$ at the serialized completion transition.
 \item Every release agrees with the authorized projection of
 $\mathcal F(D_e,q,\mathcal I)$ at its serialized release head $\Head_e$.
 \item An edit preserves $\Base$ and all prior charges and consumption marks.
 Reconstruction additionally preserves $\Head$, revokes old pending
 originals, and transfers only actually held unused preparations, made ready
 by a matching acknowledgment from the new executor.
\end{enumerate}
Thus retention preserves computational validity without replenishing private
authority. Public cloning alone creates no unused private preparation.
\end{definition}

\paragraph{Why the clauses must be coupled}
Consider a modified issuance rule that reuses $(z,M_z,e_z)$ for queries
$q$ and $q'$. Equation~\eqref{eq:query} then gives
$x_z(q)-x_z(q')=\pi_r(\mu_q-\mu_{q'})\bmod Q$.
Both evaluations can be correctly authenticated, yet the query messages
are no longer hidden by independent phases. Similarly, correct evaluation
of an earlier database version does not establish current release authority,
and restoring a pending entry without its consumption mark permits a second
private attempt on an identical authenticated result. These examples concern
altered transitions of our protocol. The invariant excludes them jointly,
rather than treating evaluation validity as sufficient release authority.

\begin{theorem}[Exact current-output correctness]
\label{thm:correctness}
Assume an honest source, honest input policies, the execution and owner model of
Section~\ref{sec:model}, and the source bounds of Lemma~\ref{lem:phase} with
$2B_\nu<Q$ for every allowed terminal type $\nu$.
Assume also that the terminal codec decodes every authorized result exactly.
Before an authentication or binding failure, every released
output equals the authorized projection of $\mathcal F(D_e,q,\mathcal I)$
at its release head.
Private completion is at most once per original, and neither data edits nor
worker reconstruction can restore an attempted capability.
\end{theorem}
\begin{proof}
Setup fixes the honest base and empty history and ledgers.
Algorithms~\ref{alg:issue}--\ref{alg:admit} and Lemma~\ref{lem:oneuse} preserve
exclusive custody and monotone consumption. Edits mark older candidates stale,
and reconstruction revokes old generation authority.

An admitted result is the complete prescribed computation for its bound
original. Lemmas~\ref{lem:affine} and~\ref{lem:phase}, with exact decoding, give
$\Finish(y)=v_0(q)$ by Equation~\eqref{eq:sign}. Lemma~\ref{lem:correction}
therefore makes Algorithm~\ref{alg:admit}'s corrected vector exact at $\Head_e$.
$\Top$ supplies the bound identifiers with row-index tie breaking. Exact cached
scores yield the same answer. The lock serializes admission, correction, cache
selection, edits and release. An edit after issuance stales its candidate.
A release before an edit linearizes at the earlier head. Stale or superseded requests and
failures release nothing. Induction proves exactness, and
Lemma~\ref{lem:oneuse} gives the capability claim.
\end{proof}

\section{Adaptive Security}
\label{sec:security}
\subsection{Games and Primitive Assumptions}
The integrity adversary controls the public host and packet delivery, and
observes the honest owner's public admission outcomes.
$\mathsf{Bad}$ occurs if an unauthorized value reaches private completion,
an original invokes private completion more than once, or a released answer
is incorrect or stale for its linearization head.
The finite multi-session EUF-CMA and binding-collision losses sum to
\begin{equation}
 \varepsilon_{\rm auth}=
 \varepsilon_{\rm ownerSig}+\varepsilon_{\rm workerSig}
                         +\varepsilon_{\rm hash}
 \label{eq:auth}
\end{equation}
The hash term covers complete originals, assets, and signed payloads.
Signatures need ordinary unforgeability, not strong unforgeability:
a new signature on an already authorized identical message causes no false
binding.

For privacy, the environment supplies two causally $\mathcal L$-compatible
adaptive policies. In world $b$, the source and owner use policy $b$.
The host sees only the declared public view and outputs a bit.
Its public-view distinguishing advantage is
\begin{equation}
 \Adv_{\rm priv}(\mathcal A)=
 \left|\Pr[\mathcal A(\mathsf{View}_0)=1]
       -\Pr[\mathcal A(\mathsf{View}_1)=1]\right|.
 \label{eq:privacygame}
\end{equation}
Both policies use exact private ideal outputs for subsequent decisions and
remain independent of secret-key state and hidden encryption randomness.

For recipient $v$, let $\varepsilon_v(M_v,\tau_v)$ bound the decisional
Ring-LWE advantage between at most $M_v$ ordinary samples
\[
 (a,-as_v+e)\quad\hbox{and}\quad(a,u),
\]
where $a,u$ are independent uniform elements of $R_Q$, $s_v$ has the
specified ternary distribution, and $e$ has independent $\CBD_\eta$ coefficients.
$\tau_v$ includes the reduction's complete simulation time.
The invertible map $(a,b)\mapsto(ta,tb)$ gives the exact scaled-error distribution
$(a,-as_v+te)$ used in Equation~\eqref{eq:parent}.
After scaling, the sample coordinates supply the mask and zero-message first
component. Adding $\mu$ to the latter gives the ciphertext.
This assumption is for the actual ring and distributions. A concrete
security estimate is a separate parameter-selection question.

Public expansion uses a domain-separated ideal classical extendable-output
function (XOF) with independent
uniform byte streams per input. Its low-bit rejection sampler produces
uniform coefficients below $Q$. Let $\kappa$ be the bit length of fresh seeds
and random protocol names, $P_s$ count all fresh parent seeds, and $h$ count
adversarial XOF input queries. Let $V$ count the remaining fresh random names.
Coupling to ideal interfaces with distinct identifiers gives
\begin{align}
 \delta_X&\le \frac{P_sh}{2^{\kappa}}+
                 \frac{P_s(P_s-1)}{2^{\kappa+1}},\label{eq:xofbad}\\
 \delta_{\rm names}&\le \frac{V(V-1)}{2^{\kappa+1}}.\label{eq:namebad}
\end{align}
Failed and unused reservations count. Identical republishing does not.
For a strict fair-bit polynomial-time reduction, cap each rejection stage
at $\rho$ candidates. With $S_{\rm all}=\sum_vM_v$ samples and $R_s$ fresh
role-secret draws,
\begin{equation}
 \delta_{\rm cut}\le C_{\rm samp}2^{-\rho},
 \qquad C_{\rm samp}=N(2S_{\rm all}+P_s+R_s).
 \label{eq:cut}
\end{equation}
This conservative count includes uniform sample components, conditional XOF
streams, and ternary draws. CBD draws have fixed length.
Appendix~\ref{app:sampling} details the exact stream distributions and sampling cutoff rules.

\subsection{Integrity and Ideal Private Completion}
\begin{lemma}[Admission and ideal private completion]
\label{lem:virtual}
Assume the correctness hypotheses of Theorem~\ref{thm:correctness}, including
$2B_\nu<Q$ for every allowed terminal type and pointwise exact terminal
decoding, together with the authentication and binding games above. Then
\[
 \Pr[\mathsf{Bad}]\le\varepsilon_{\rm auth}+\delta_{\rm names}.
\]
Before such failure, substituting exact $v_0(q)$ for authorized $\Finish(y)$
calls, with Algorithm~\ref{alg:admit}'s correction and release unchanged,
preserves the declared distribution of public traces.
\end{lemma}
\begin{proof}
Accepting an unauthorized owner request gives a signature forgery or binding
collision. The confined executor signs only complete prescribed results, so
an accepted nonprescribed result implies a worker forgery or binding collision. Extent, domain and
pending nonce exclude partial-result and cross-original substitution.
Theorem~\ref{thm:correctness} and Lemma~\ref{lem:oneuse} preclude duplicate
completion and incorrect or stale release. Name coupling adds $\delta_{\rm names}$.

Source bounds and exact decoding give $v_0(q)$. Lemma~\ref{lem:correction}
gives the ideal current answer.
Public rejection and stale/cache finalization inspect only public bindings and
lifecycle state, without private decoding. Represent delayed preparation by
an opaque parent-sample handle, exposing only $a_z$ until unique issue, then
adding the honest message to its hidden first component and exposing the projection.
Unsent error coordinates
are independent and integrate out. Policies cannot inspect masks or errors, so
early and issue-time sampling have identical observations. Induction preserves
public state, pending bindings and private ideal answers, including adaptive
decisions. Later ciphertext hybrids use this ideal owner without decrypting
changed ciphertexts.
\end{proof}

Invalid host bytes never reach a secret-dependent decoder. Prescribed honest
results meet the decoding bound for every allowed secret and error. Rejection
therefore depends on public validity and authority, not attacker-selected phase
tests. This reaction-attack defense relies on the execution and owner premises.
It does not make BGV generally CCA-secure.

\subsection{Descending Recipient Reduction}
\begin{theorem}[Bounded adaptive input privacy]
\label{thm:privacy}
For the acyclic independent-recipient protocol, assume the correctness
hypotheses of Theorem~\ref{thm:correctness}, including $2B_\nu<Q$ for every
allowed terminal type and pointwise exact terminal decoding.
Under Section~\ref{sec:model}'s execution premises and the games above,
\begin{equation}
\begin{aligned}
 \Adv_{\rm priv}(\mathcal A)\le 2\biggl(
       &\varepsilon_{\rm auth}+\sum_v\varepsilon_v(M_v,\tau_v)\\
       &+\delta_X+\delta_{\rm names}+\delta_{\rm cut}\biggr).
\end{aligned}
 \label{eq:privacybound}
\end{equation}
\end{theorem}
\begin{proof}
In each world, stop on authentication/binding failures and couple names and
bounded samplers at the displayed cost.
Apply Lemma~\ref{lem:virtual} before changing any ciphertext distribution.
All completions now use ideal private answers.

For descending replacement, let $\mathsf{Hyb}_c$ keep recipients $v\le c$ real
and replace parent and hint first components addressed to $v>c$ by independent
uniform elements of $R_Q$.
For the adjacent reduction at $c$, draw all other secrets independently and
obtain $c$'s samples from its ordinary Ring-LWE oracle, followed by the
invertible $t$ map. Every still-real target incoming to $c$ depends only on
known lower secrets, so the reduction adds it to the oracle's masked component.
All outgoing hints involving the unknown $s_c$ have higher recipients and
are already uniform. No unknown target is evaluated.
Zero parents and honest gallery messages are simulated by the same operation.

For a query-preparation reservation, expose the oracle's uniform $a$ using a fresh
programmed XOF input and keep its zero-message ciphertext component hidden.
Programming samples the \emph{complete} conditional byte stream, including
rejected candidates, discarded high bits, and later suffix, so the
real-oracle distribution equals that of the actual seeded parent.
Prequeries and repeated fresh inputs cost $\delta_X$.
Retained preparations use only $a$ and public assets.
At unique issue the policy fixes its message without inspecting that component.
In the uniform-oracle case, translation by this adaptive message preserves
conditional uniformity. Each projected row has the same number of preimages.
The real-oracle case gives exactly the projected parent marginal.

Public arithmetic, terminal conversion, hashes, and signatures are
postprocessing of these simulated samples.
Namespaces are computed from their actual bytes in each hybrid.
The same ideal owner and causal policy interfaces continue all later events.
Hence the adjacent gap is at most $\varepsilon_c(M_c,\tau_c)$.

When all recipients are uniform, every exposed first component is independent of
the data/query messages. A common simulator maintains public ledgers,
generation/head names, bindings, and public evaluations from the declared
actions and leakage. Ideal owners receive their respective exact private
answers. Compatibility gives the same distribution of the next public transition.
The simulator never decodes uniformized results.
Summing adjacent gaps bounds each world's distance to this common view.
The triangle inequality gives Equation~\eqref{eq:privacybound}.
\end{proof}

The two structural requirements are essential to this reduction.
Secret-dependent same-key targets are not covered by the descending-recipient
argument and require a separate key-dependent-message security argument.
Issuing a reserved parent twice lets two chosen messages share
one mask/error, revealing their difference. Independent recipients and
one-use issuance remove these obstacles in the stated construction.

\subsection{Key Lifetimes and Composition}
For gadget width $k=\ceil{\log_\beta Q}$, fresh gallery parents $A_g$, and
query-preparation reservations $T_q$, the sequential per-recipient sample budgets are
\begin{align}
 M_0&=1+T_q,&M_v&=1+2k\quad(0<v<L),\notag\\
 M_L&=1+2k+A_g,&M_f&=
 \begin{cases}1+2k&\text{pair},\\1+k&\text{mixed}.\end{cases}
 \label{eq:lifetime}
\end{align}
Tensor has no final role. With $g_n$ gallery groups, one encryption per
feature and group gives $A_g=g_nd$ at initial setup. Across $J_g$ generations
reusing the graph and admitting at most $T_{\rm gen}$ fresh reservations per
generation, use $T_q\le J_gT_{\rm gen}$ and count every fresh gallery
reencryption. Reconstruction that only republishes existing encrypted assets
and truthful local correction create no new HE parent by themselves.
The budget counts abandoned and unused reservations as well as issued queries.

\paragraph{Security-parameter interpretation}
Theorem~\ref{thm:privacy} is a finite reduction, stated for arbitrary admissible
parameters and lifetime budgets. For a family indexed by $\lambda$ with
polynomially many recipients and polynomial lifetime budgets, it establishes
negligible privacy advantage if the actual Ring-LWE sample games and the
authentication primitives have negligible advantages, the aggregate
coupling terms are negligible, and complete simulation time is polynomial.
The Ring-LWE assumption must hold for the chosen ring, modulus, secret and error
distributions, and sample budgets. No numerical hardness estimate is needed to state
this conditional result.
For example, choosing
$\rho\ge\lambda+\ceil{\log_2 C_{\rm samp}}$ makes
$\delta_{\rm cut}\le2^{-\lambda}$.
The seed/name length $\kappa$ must likewise accommodate the lifetime counts
in Equations~\eqref{eq:xofbad}--\eqref{eq:namebad}. These statistical budgets
do not assign a computational security level to a fixed Ring-LWE tuple.

A deployment with cryptographic confidential channels or an encrypted
plaintext-cache transfer adds their multi-session distinguishing and
authentication terms. Faithful enrollment and nonrollback owner state are
explicit execution assumptions, rather than consequences of Ring-LWE.

\section{Key-Capped Terminal Representation}
\label{sec:compression}
\subsection{Pointwise Rounding and Conservative Fallback}
We round complete results from computation modulus $q_c$ to wire modulus $p$,
retaining $t$ and using $\ceil{\log_2p}$ bits per coefficient.
Let $B$ bound the complete integer phase, including scores and noise.
The baseline uses $(q_c,B)=(Q,B_\nu)$ from Equation~\eqref{eq:bounds}.
Section~\ref{sec:retainedjoin} uses $q_c=P$ with its own complete-phase bound.
Let $\boldsymbol w$ be the decryption multiplier vector from
Equation~\eqref{eq:types}. Its aggregate coefficient norm
$\Lambda=\sum_i\norm{w_i}_1$, summed over terminal components, measures
rounding-error amplification.

Choose an odd wire modulus $1<p<q_c$ with $p\equiv q_c\bmod t$.
For a canonical coefficient $0\le c<q_c$, select the nearest integer to $(p/q_c)c$
in $c+t\mathbb Z$, rounding halfway ties upward:
\begin{equation}
 r_p(c)=t\left\lfloor
 \frac{2(pc-q_c(c\bmod t))+q_ct}{2q_ct}\right\rfloor+c\bmod t.
 \label{eq:round}
\end{equation}
This BGV terminal modulus switching~\cite{bgv} preserves
$r_p(c)\equiv c\bmod t$. Eligible integers are spaced $t$ apart, giving
$|r_p(c)-(p/q_c)c|\le t/2$. The wire stores $r_p(c)\bmod p$.

\begin{lemma}[Whole-phase terminal correctness]
\label{lem:round}
Suppose an authorized source terminal has centered phase lift $v$ with
$\norm{v}_\infty\le B<q_c/2$.
If
\begin{equation}
 B'=\ceil{pB/q_c}+\ceil{t\Lambda/2}<p/2,
 \label{eq:terminalbound}
\end{equation}
then the reduced terminal has the same plaintext at every coefficient.
\end{lemma}
\begin{proof}
Let $\boldsymbol C$ be the integer canonical component lifts. Then
$\sum_iC_iw_i=v+q_ck_0$, with carry polynomial $k_0\in R$.
Set $\widetilde C_i=r_p(C_i)$ coefficientwise and
$v'=\sum_i\widetilde C_iw_i-pk_0$.
The rounding errors have coefficient bound $t/2$, so
$\norm{v'}_\infty\le pB/q_c+t\Lambda/2$.
The canonical wire phase is congruent to $v'$ modulo $p$.
Equation~\eqref{eq:terminalbound} makes $v'$ its unique centered lift.
Since each rounded lift preserves its residue modulo $t$ and $p\equiv q_c\pmod t$,
$v'\equiv v\bmod t$. Individual wire components need not preserve that
residue. The argument applies to the complete phase including its carries.
\end{proof}

The guard prevents wraparound modulo $p$. Appendix~\ref{app:codec} illustrates
both terms of $B'$ and distinguishes rounded lifts from stored residues.

At setup, privately compare $\Lambda$ with the public type-specific cap
$\Lambda_\nu^\star$ (Appendix~\ref{app:keycaps}). Select the smaller wire profile
if it passes, otherwise the worst-case fallback. Bind the type and profile in
$\Base$ and enrollment, without resampling the key. Both profiles satisfy
Equation~\eqref{eq:terminalbound} for every authorized query at their admitted
keys, without independence assumptions on ciphertexts or rounding errors.
A failed cap selects a larger response without a decryption failure.

A pair can instead select a profile per response using a \emph{public residual
certificate}, a sufficient rounding-error check under an admitted key-support
bound, not an execution-authentication proof. Enrollment fixes the permitted
profiles, selection rule and support cap.
For canonical component polynomials $c_i$, form the residual bounds
$\rho_i=\ceil{|r_p(c_i)-(p/q_c)c_i|}$ coefficientwise.
Let $\TopSum_K$ sum the $K$ largest entries, where
$K\in\{0,\ldots,N\}$ is the registered support cap. At a ternary recipient with
$\norm{s_f}_1\le K$, the sufficient guard is
\[
 2\bigl(\ceil{pB/q_c}+\norm{\rho_0}_\infty+
            \TopSum_K(\rho_1)\bigr)<p.
\]
Negacyclic shifts use at most $K$ residual entries in each phase coefficient.
Appendix~\ref{app:residual} proves the certificate. A conservative fallback
from Lemma~\ref{lem:round} makes the registered selector a total function on
canonical inputs of the enrolled shape. One profile must pass every group
and bind the complete signed body, without changing $\Base$.
Malformed frames are rejected publicly.
Totality also holds on uniformized hybrid inputs, without asserting their
plaintext correctness.

\subsection{Profile Exposure and Privacy}
For pair terminals, $\Lambda=1+\norm{s_f}_1$. Choose a public threshold
$K_s>2N/3$. Since $\norm{s_f}_1$ is $\operatorname{Bin}(N,2/3)$,
Hoeffding bounds the fallback probability by
\[
 \delta_0\le\exp\bigl(-2(K_s-2N/3)^2/N\bigr).
\]
Appendix~\ref{app:keycaps} gives the corresponding mixed/tensor bounds.
The bound concerns public profile exposure, not computational Ring-LWE
hardness. A profile predicate is evaluated once without resampling the key.

\begin{corollary}[Profile selection and adaptive privacy]
\label{cor:cap}
Predeclare the terminal types and a maximum fresh-key lifetime. Let $\mathcal K$ index
the allowed fresh family/type pairs, with terminal-type index $\nu(\xi)$ for
pair $\xi$. Bound lifetime profile exposure by
$\Delta_{\rm cap}=\sum_{\xi\in\mathcal K}\delta_{\nu(\xi)}$.
Count each distinct pair once, including all allowed pairs when their actual
exposure is adaptive. Reusing an already exposed pair adds no event.
The key-capped protocol retains the integrity bound of
Lemma~\ref{lem:virtual} and satisfies
\begin{equation}
\begin{aligned}
 \Adv_{\rm priv}\le 2\biggl(
       &\varepsilon_{\rm auth}+\sum_v\varepsilon_v\\
       &+\delta_X+\delta_{\rm names}
                         +\delta_{\rm cut}+\Delta_{\rm cap}\biggr).
\end{aligned}
 \label{eq:cappedprivacy}
\end{equation}
\end{corollary}
\begin{proof}
Both actual profiles decode exactly by Lemma~\ref{lem:round}, so the integrity
argument is unchanged. For privacy, we first idealize authorized private completion
and couple actual selection to a virtual always-capped public profile with
ideal private completion. Outside any exposed bad-cap event the public
trace is identical, so the one-world statistical distance is at most
$\Delta_{\rm cap}$. The virtual owner uses ideal answers even on keys that
would require fallback, without decoding their reduced terminals.
We apply the recipient-hybrid argument from Theorem~\ref{thm:privacy}'s proof
to this virtual protocol with its unchanged, unconditioned secret distributions.
The reduction does not condition a Ring-LWE challenge on passing a secret predicate.
All compression is public postprocessing. The two-world triangle inequality
adds $2\Delta_{\rm cap}$. The union bound covers shared-key correlations.
\end{proof}

\begin{theorem}[Retention-preserving security composition]
\label{thm:retentioncomposition}
Fix arithmetic parameters, terminal types and wire profiles satisfying
Theorem~\ref{thm:correctness}'s correctness hypotheses, and finite lifetime
budgets. Under the execution model and primitive assumptions of
Theorem~\ref{thm:privacy} and Corollary~\ref{cor:cap}, the key-capped protocol
maintains Definition~\ref{def:retention}'s invariant except with probability
\begin{equation}
 \Pr[\exists\tau:\neg\mathsf{Inv}_{\rm ret}(\tau)]
 \le\varepsilon_{\rm auth}+\delta_{\rm names}.
 \label{eq:retentionfailure}
\end{equation}
For causally $\mathcal L$-compatible adaptive policies, its privacy advantage
is bounded by Equation~\eqref{eq:cappedprivacy} over the same lifetime.
Retained recomputation, edits, and worker replacement introduce no additional
term, provided all fresh samples, signing generations, and exposed family/type
pairs are counted in the stated games and budgets.
\end{theorem}
\begin{proof}
Outside the authentication, binding, and name-collision events of
Lemma~\ref{lem:virtual}, Lemma~\ref{lem:oneuse} proves the first invariant
clause. Whole-result binding and consumption before completion prove the
second. Theorem~\ref{thm:correctness}, Lemma~\ref{lem:correction}, and the
pointwise terminal bound prove the third under either selected profile.
The specified edit and reconstruction transitions prove the fourth, including
revocation of old pending originals and the transfer of unused objects rather
than reconstruction of private capability from public records.

These clauses justify ideal private completion before recipient hybrids.
Unique issuance exposes each delayed parent first component once, and exact current
answers drive later adaptive policies. Recomputing from public assets is
postprocessing, not a fresh sample. Corollary~\ref{cor:cap} charges selection
once per allowed family/type pair. Retention and reconstruction therefore
add no loss.
\end{proof}

\section{Retained Joins at a Smaller Modulus}
\label{sec:retainedjoin}
Expansion at $Q$ precedes a retained ciphertext--ciphertext join at $P<Q$.
Ordinary BGV modulus switching~\cite{bgv,bgvdependencies} can incur rounding
loss that prevents a useful smaller $P$. Our compiler invariant permits a
simultaneous bound valid after adaptive query and reservation selection,
while preserving the original's completion authority.

\subsection{Split Construction and Retained Equality}
For the split join, choose $1<P<Q$ with integer ratio $a=Q/P>1$ odd,
$a\equiv1\pmod t$, and $\gcd(t,Q)=1$. Thus
$P\equiv Q\pmod t$. Write $x=x_z(q)$ and
$C_j(x,z)=\Expand(x,a_z)_j$ for a complete source leaf.
Denote by $r_{Q\to P}$ the unreduced integer rounding map
of Equation~\eqref{eq:round}, extended to integer inputs. It preserves
residues modulo $t$, has residual magnitude at most $t/2$, and satisfies
$r_{Q\to P}(c+Qk)=r_{Q\to P}(c)+Pk$ for integers $c,k$.
Lemma~\ref{lem:affine} gives
\[
 C_j(x,z)=C_j(0,z)+(\delta_j,0)\pmod Q,
 \qquad \delta_j=\overline{Nx_j}^{\,Q},
\]
where $\overline{u}^{\,Q}\in\{0,\ldots,Q-1\}$ is the canonical residue
and $\delta_j$ is inserted at coefficient zero. Define on canonical lifts,
coefficientwise,
\begin{align}
 U_j(z)&=(r_{Q\to P}(C_{j0}(0,z)),r_{Q\to P}(C_{j1}(0,z)))\bmod P,\notag\\
 \gamma_j(x)&=r_{Q\to P}(\delta_j)\bmod P,\notag\\
 L_j(x,z)&=U_j(z)+(\gamma_j(x),0)\bmod P.
 \label{eq:joinsplit}
\end{align}
Here $U_j$ is the retained seed-only pair, $\gamma_j$ is the late rounded
scalar, and $L_j$ is the completed query leaf. This enrolled circuit generally
differs from rounding $C_j(x,z)$ as a whole, since rounding is not additive.
The incomplete $U_j$ is never privately decoded and has no fresh-phase bound.
The late first component cancels the mask to yield the bounded complete phase.

Encrypt each gallery feature freshly at $P$ under the same integer
secret polynomial $s_g$, with fresh randomness, giving $B_j\in R_P^2$.
Final hints also use modulus $P$, with targets $s_g,s_g^2$ and recipient $s_f$.
The seed-only database join is
$T_z=\sum_{j<d}U_j(z)\otimes B_j$. Its completed value obeys the exact identity
\begin{equation}
 \sum_{j<d}L_j(x,z)\otimes B_j
 =T_z+\left(\sum_{j<d}\gamma_jB_{j0},\sum_{j<d}\gamma_jB_{j1},0\right).
 \label{eq:joinretained}
\end{equation}
The unchanged quadratic component is switchable during preparation. For a pair
terminal, the linear component is switched after the late update. Both moduli,
gadget radices, graph, scalar map, terminal type and codec are enrolled in
$\Base$. An original authorizes this complete circuit once.

\subsection{Query-Independent Rounding Family}
A \emph{catalog} bundles authenticated public assets for a fresh key family,
including hints and the encrypted index. An admitted catalog fixes a base
$\Base$. Worker reconstruction reuses it under a new $\Generation$.
The owner's internal lifecycle
controller fixes positive integer limits $C$ on catalog draws and $H$ on
query-preparation reservations per catalog, including failed, unused and retired draws.
The source fixes the expansion graph and join profile, then samples $s_g$ with
independent uniform ternary coefficients, independently of its pre-draw history.
Here $g$ indexes the graph's gallery recipient, equal to $L$ in the sequential
graph. Let $\mathcal A$ collect all hint second components, sampled independently
of the secrets.
The implementation also supports a \emph{blocked} graph, grouping one or two
adjacent expansion levels under one new recipient while preserving the sequential
extraction map (Appendix~\ref{app:joingraph}). Both compilers decompose only the
input second component. The recurrence
uses these digits, hint second components and public signed automorphisms or
monomial maps, never first components, errors or query bits. Thus, for every
honestly reserved seed $z_h$, its leaf's second component is
\begin{equation}
 C_{j1}^{(h)}:=C_{j1}(0,z_h)=f_j(\mathcal A,z_h,\mathsf{graph}),
 \label{eq:joinnoninterference}
\end{equation}
Here $f_j$ is the deterministic public recurrence, and the mask family
$(\mathcal A,z_1,\ldots,z_H)$ is independent of $s_g$.
Seed expansion uses only its fixed domain, seed, public dimensions and layout,
modulus and counter, never a secret-dependent catalog hash.

There are no free resets, peer-selected seeds or
searches using $s_g$. Final-recipient admission does not filter $s_g$.
A graph or join-modulus change requires a fresh charged catalog and $s_g$.
Enrolled wire-profile selection leaves this circuit unchanged. These constrain
honest sampling, not secret-coefficient independence after public observations
or slot selection. The \emph{reservoir} is the conceptual family of all $H$
permitted reservation seeds per catalog, sampled independently of $s_g$.
A run need only generate its
consumed prefix.
Unlike Section~\ref{sec:compression}'s larger-profile fallback, a failed setup
cap on the final recipient may abort its charged catalog without refund.
This admission check precedes any per-response residual selection.
We couple admission to a virtual source that always admits the unconditioned final
recipient and uses ideal completion. Transcripts agree outside failed caps.
Counting aborted draws, at most $C\delta_0$ per world enters $\Delta_{\rm cap}$.
This extends Corollary~\ref{cor:cap}'s profile coupling without oracle
conditioning or within-catalog retries.
We recompute Equation~\eqref{eq:lifetime}'s sample counts for the enrolled graph,
charging every fresh zero parent, hint column, gallery parent and reservation
to its recipient. Each direct-$P$ sample requires a distinct source-$Q$ oracle call.

\begin{lemma}[Simultaneous rounding under adaptive selection]
\label{lem:joinreservoir}
For reservation $h$ and feature leaf $j<d$, suppress $(h,j)$ in the residual
notation and let
$\epsilon_i=r_{Q\to P}(C_{j1,i}^{(h)})-(P/Q)C_{j1,i}^{(h)}$,
$\rho_i=\ceil{|\epsilon_i|}$, $S_2=\sum_i\rho_i^2$ and $M=\max_i\rho_i$.
Thus $M$ bounds the largest residual and $S_2$ bounds their squared weight
sum. For tail parameter $\ell_J>0$, bound convolution magnitudes by
\begin{equation}
 Z=\ceil{\frac{M\ell_J+\sqrt{M^2\ell_J^2+12\ell_JS_2}}3}.
 \label{eq:joinbernstein}
\end{equation}
Under the source sampling assumptions above, every coefficient of the error polynomial
$\epsilon*s_g$ for every honestly reserved feature leaf is bounded by its
corresponding $Z$, except with probability at most
\begin{equation}
 \delta_J=2CHdN e^{-\ell_J}.
 \label{eq:joinfailure}
\end{equation}
Here $*$ denotes negacyclic convolution. The event covers adaptive binary
queries and selection of reserved slots.
\end{lemma}
\begin{proof}
For one catalog, we condition on its pre-draw history, fixed graph and
independent mask family. A coefficient of $\epsilon*s_g$ is a signed
sum of independent mean-zero ternary variables with fixed weights
$\epsilon_i$. Its variance is at most $2S_2/3$ and each summand has
magnitude at most $M$. Bernstein's two-sided bound is
\[
 2\exp\left(-\frac{Z^2}{4S_2/3+2MZ/3}\right)\le2e^{-\ell_J}.
\]
For $M=0$ the coefficient is zero. We apply a union bound over reservations,
features and coefficients, remove conditioning, then sum unconditional catalog
bounds. We do not condition these bounds on later secret-dependent choices.
Only the $d$ join leaves contribute. Coefficients and leaves need not
be independent. Queries change only first components and all honest slots are
covered, so adaptive selection adds no failure event and requires no assumption
about the resulting conditional secret distribution.
\end{proof}

For a target $\lambda>0$, choosing
$\ell_J\ge\lambda\ln2+\ln(2CHdN)$ gives $\delta_J\le2^{-\lambda}$.
Dependence between reused secrets and residuals can invalidate average-case
noise estimates~\cite{noisecritique}. The property used here is the explicit
recurrence~\eqref{eq:joinnoninterference} and finite honest family. Bernstein
and the union bound themselves are standard.

\subsection{Complete Phase and Security Composition}
Let $\beta_Q,\beta_P\ge2$ be the source and final gadget radices.
The selected query coefficient has phase bound
$F_o=\min(\alpha,t-\alpha)+t\eta$ from Equation~\eqref{eq:query}.
Put $k_Q=\ceil{\log_{\beta_Q}Q}$ and $S_Q=tN\eta k_Q(\beta_Q-1)$ for one
source switch. A two-level block maps an incoming switching-error bound $V$
to $4(V+S_Q)$, no larger than the sequential $4V+6S_Q$.
It preserves the extraction operator, so complete source leaves admit
$E_Q=NF_o+2(N-1)S_Q$ in either graph. Appendix~\ref{app:joingraph} gives the maps.
Let $Z_*$ be an enrolled upper bound on every $Z$ in the finite family,
including as-yet undrawn reservations, not merely an observed maximum.
$M\le\ceil{t/2}$ and $S_2\le N\ceil{t/2}^2$ give a uniform choice.
The final-switch error bound is
$S_P=tN\eta\ceil{\log_{\beta_P}P}(\beta_P-1)$.
Define the complete split-leaf and joined-pair bounds by
\begin{equation}
 E_J=\ceil{PE_Q/Q}+t+Z_*,\qquad
 B_J=Nd(1+t\eta)E_J+2S_P.
 \label{eq:joincompletebound}
\end{equation}
The leaf bound adds scaled source phase, two first-component rounding errors
and the secret-weighted residual. Joining $d$ gallery columns and switching
two tensor components gives $B_J$.

The final wire modulus is $p<P$, distinct from the join modulus.
Section~\ref{sec:compression} applies with $q_c=P$ and complete bound $B_J$:
for example, $\ceil{pB_J/P}+\ceil{t\Lambda/2}<p/2$ suffices.
A registered public-residual selector checks every complete group, binds one
profile and whole extent before signing, and has a total conservative fallback on canonical inputs. For support cap
$K$, registering $p_f$ with
\[
 \ceil{p_fB_J/P}+(K+1)\ceil{t/2}<p_f/2
\]
ensures totality by the residual guard of Section~\ref{sec:compression},
even on canonical hybrid inputs. It does not assert that hybrid phases obey
$B_J$. Selection does not consult private decoding success.

\begin{theorem}[Retention-preserving split join]
\label{thm:joincomposition}
For the pair-terminal split construction, assume the ring and extraction
conditions of Section~\ref{sec:arithmetic} and the source sampling assumptions of
Lemma~\ref{lem:joinreservoir}. Let $E_Q$ be a valid complete source-leaf
phase bound and $Z_*$ an enrolled uniform upper bound on
Equation~\eqref{eq:joinbernstein} over all allowed reservations.
For $B_J$ in Equation~\eqref{eq:joincompletebound}, assume $2B_J<P$, enrolled
terminal conversion exact for phases bounded by $B_J$, and the custody,
prescribed-execution, truthful-history, monotone-consumption and observation
premises of Theorem~\ref{thm:retentioncomposition}. Replacing its expansion
and join by Equations~\eqref{eq:joinsplit}--\eqref{eq:joinretained}
preserves its authorized current-output invariant except with additional
probability $\delta_J$. Its adaptive privacy bound gains at most
$2\delta_J$. All sample and exposure budgets are recomputed for the actual
enrolled graph and charged lifetime, using ordinary Ring-LWE games at $Q$.
\end{theorem}
\begin{proof}
For integer component lifts, the complete source identity is
$C_{j0}(0,z)+\delta_j+C_{j1}(0,z)s_g=v+Qk$ with carry $k\in R$ and
$\norm{v}_\infty\le E_Q$. Scale it by $P/Q$. The split phase differs by
two first-component residuals of magnitude at most $t/2$ and
$\epsilon*s_g$, giving $E_J$ outside the lemma's event. Subtracting $Pk$
and using $P\equiv Q\pmod t$ preserves its plaintext. Gallery phases have
bound $1+t\eta$. Their join and two final switches give $B_J$.
The strict guard makes centering unique. The enrolled terminal certificate,
telescoping edit repair and serialized release give exact current answers.
Spend-before-completion and revocation preserve one-use authority.

In each privacy world, first couple private completion to its exact
authorized ideal output, paying $\delta_J$. Then apply the descending
recipient hybrids of Theorem~\ref{thm:privacy}. The public terminal selector
is total, including fallback, on each hybrid's own canonical inputs. No statistical
phase claim about hybrid ciphertexts is needed. Scaling an ordinary $Q$
sample by the invertible $t$ map gives the required scaled-error distribution.
Reducing a fresh such sample modulo $P\mid Q$ gives exactly a fresh direct
$P$ sample under the same integer secret. Uniform samples remain
uniform with equal numbers of preimages. Every direct $P$ sample uses a separate
$Q$ oracle call, distinct from exposed $Q$ samples. Earlier-recipient
target shifts commute with reduction, so the mixed-modulus graph introduces
no key cycle. The two-world inequality adds $2\delta_J$ to the existing
authentication, naming, sampler, seeded-mask and final-admission terms.
Concrete randomness and XOF substitutions retain their separate obligations.
\end{proof}

\section{Implementation and Evaluation}
\label{sec:evaluation}
\subsection{Construction Profiles and Timing Boundary}
The prototype implements the BGV arithmetic and retained construction in
custom Python/C++ code, including independent-recipient switching, native
completion, and authenticated terminal encoding. A \emph{construction profile}
fixes arithmetic, retained state, wire-profile policy and completion path. We compare
three: Q120 evaluates at a 120-bit source modulus, P64 joins at a 64-bit
modulus using a deterministic worst-case rounding bound, and P60 joins at a
60-bit modulus using the simultaneous statistical bound of
Section~\ref{sec:retainedjoin}. All use blocked independent-recipient expansion and
the same dimensions and packing layout: $N=16384$, $d=512$, two groups, and
32,768 ordered binary records. Q120 and P64 use a fixed 24-bit terminal
codec. P60 uses a 116-bit source modulus, a 60-bit join modulus,
$t=1031$, error parameter $\eta=21$, source radix $2^{30}$, and final
radix $2^{20}$. Its public residual selector chooses among prescribed
23-, 24-, and 25-bit representations. The honest family declares
$C=10$ catalogs and $H=16$ reservation seeds per catalog, counting failed and
unused draws. These budgets are premises on honest construction, rather
than a limit on arbitrary adversarial processes.

Each profile creates a fresh source, transfers and verifies its complete signed
seeded catalog, imports the assets and prepares two independent slots consumed
by distinct queries. The \emph{paid} time is wall-clock time from before source
creation through finalization and cleanup of native handles, transports and
threads. It covers this two-query workload, not the permitted key lifetime.
Setup has five framed transfers:
the signed publication, public keys, switching hints, seeded index, and
ordered identifiers. Input-fixture acquisition and independent plaintext
correctness checks are outside the measured interval. No profile receives
cloned source custody or free preparations.

Measurements use a Linux Ryzen 7 5800X host with eight physical cores and
sixteen logical CPUs, including asset and result exchange over unpaced local
Unix sockets. A \emph{cohort} runs all three profiles once. One calibration
freezes the execution policy before three held-out cohorts with rotated profile
order.

\subsection{Complete Costs and Representation}
\begin{table}[t]
\makeatletter
\long\def\@makecaption#1#2{%
  \par\vskip\abovecaptionskip
  {\normalfont\footnotesize\raggedright\textbf{#1.} #2\par}}
\makeatother
\renewcommand{\tablename}{Table}
\centering\small\setlength{\tabcolsep}{4pt}
\CombPaidProfileTable
\caption{Medians across three held-out cohorts. Preparation is the
total for two preparations. Public and private entries are medians of the
cohort means over two queries. Public includes evaluation and exchange.
Private includes authorization, completion, and ranking. Paid covers the
complete workload, including setup and cleanup.}
\label{tab:paid-profiles}
\end{table}

Table~\ref{tab:paid-profiles} shows that P60's reduced source construction,
import, and public evaluation costs survive its additional preparation and
private-path costs in this two-query workload. The median within-cohort
Q120/P60 paid ratio is \CombQPaidRatio{}
[\CombQPaidRatioMin{}, \CombQPaidRatioMax{}]. P64/P60 is
\CombDetPaidRatio{} [\CombDetPaidRatioMin{}, \CombDetPaidRatioMax{}].
These are three paired ratios and their full observed ranges, rather than
ratios of table medians or confidence intervals. After preparation, first-answer
latency runs from the start of query issuance to delivery of its authorized
ranked output. It is \CombQFirstMs{}\,ms, \CombDetFirstMs{}\,ms, and
\CombStatFirstMs{}\,ms for Q120, P64, and P60. The paired first-answer
ratios are \CombQFirstRatio{} and \CombDetFirstRatio{}, respectively.
Preparation remains slower than Q120, and authorization, private completion,
and ranking are slower than both controls.

Actual setup wire uploads are \CombQSetupMiB{}\,MiB,
\CombDetSetupMiB{}\,MiB, and \CombStatSetupMiB{}\,MiB.
Every observed P60 response selects 24 bits, giving the same
\CombStatBodyKiB{}\,KiB encrypted body as both controls. Its authenticated
wire profile adds 100 bytes to the framed signed response relative to both controls.
The current native gallery stores \CombStatGalleryMiB{}\,MiB of residue
payload for P60 versus \CombQGalleryMiB{}\,MiB in the two controls.
The latter extents follow their frozen 64-bit-word layouts. Peak retained
residue payload for two preparations is \CombStatRetainedMiB{}\,MiB
versus \CombQRetainedMiB{}\,MiB. These figures describe native residue storage,
excluding process overhead, rather than peak process memory or seeded wire
sizes.

Arithmetic, retained state, wire profiles and completion paths differ, so this
is not a pure modulus ablation.
A stronger capped Q120 residual codec can produce a 23-bit,
184\,KiB body, 8\,KiB smaller than the observed P60 body. It is not the
paid control here. Packing P64's two residues into 32-bit arrays could
also match P60's gallery payload. The comparison establishes neither
optimality nor amortization at other query counts, and does not show an
advantage over permitted plaintext retention. It measures CPU execution,
not WAN, GPU, or attested deployment latency. With $\ell_J=111$, the declared
reservoir gives statistical correctness bound $\delta_J<2^{-128}$.
Computational security remains conditional on the stated Ring-LWE family.

\subsection{Implementation Soundness}
Completion authenticates the whole result and profile, rejects noncanonical
encodings and durably consumes the original before decoding. Separate functional
tests cover repair across edits, edits concurrent with completion, plaintext-cache
handoff and retirement of cache-superseded results without decoding. They exercise
the baseline and Appendix~\ref{app:editcommuting}'s variant, which replaces
issue-head equality with live correction authority rather than preserving the
unchanged baseline invariant. The timed comparison instead uses a fixed head
without edits or cache races. These tests do not replace the proof's
execution-integrity and observation assumptions.

\section{Related Work}
\label{sec:related}
\subsection{Encrypted Search and Retained Preparation}
HERS~\cite{hers} searches encrypted compressed embeddings. PPMI~\cite{ppmi}
caches encrypted representations and studies query transformations and
update-related invalidation. Litchev and Ouyang~\cite{paillierhamming} accelerate
binary Hamming-distance evaluation using carry-separated Paillier encoding,
lookup encryption, reduced-exponent decoding and CUDA. Beyond encrypted
similarity and reusable representations, our functionality requires exact
current scores with row-index tie breaking under changing completion authority.

ZipPIR~\cite{zippir} moves encrypted computation offline and reconstructs server
hints after database updates without client interaction. FlashPIR~\cite{flashpir}
prepares message-independent key switching for expansion, folding and packing,
using each offline seed once. Offline affine decomposition and one-query seeds
therefore have predecessors. Our invariant specifies what survives edits or
executor replacement without renewing issuance rights, and which current owner
state authorizes completion. The smaller-modulus construction bounds the
completed split join, retaining its quadratic tensor while separately rounding
the late first-component correction.

\subsection{Evaluation Integrity and Current Completion}
Viand et al.~\cite{vfhe} analyze FHE integrity and reaction attacks. The later
vFHE framework~\cite{vfhe2024} provides integrity tools and comparisons.
Argos~\cite{argos} verifies FHE execution on commodity hardware, binding inputs
and multi-round transcripts and checking results before decryption, separately
from secret-key custody. Slalom~\cite{slalom} combines trusted hardware, masked
delegation and randomized verification. Our executor adopts this established
boundary, not a stronger attestation primitive.

BioZKFHE~\cite{biozkfhe} combines encrypted similarity, specialized BGV packing,
public verification and committee-governed release. Current authenticated data
already underlies verifiable retrieval and updates~\cite{benabbas}, IntegriDB's
query verification against a current digest~\cite{integridb}, and Hyperproofs'
dynamic vector-commitment proofs~\cite{hyperproofs}. Our construction combines
immutable-base encrypted evaluation and owner-side repair with one-use projected issuance,
consumption before private arithmetic and serialized current release.
AGATE~\cite{agate} provides a general attested-functionality framework. Our result
specifies the custody and observation obligations for this search protocol.

\subsection{Noise Bounds and the Completed Join}
BGV~\cite{bgv} supplies leveled key switching and compatible modulus conversion.
Biasioli et al.~\cite{bgvdependencies} analyze secret-weighted rounding and key
dependencies using uniform-residual and central-limit approximations under a
rounding-dominance condition. Gao and Zheng~\cite{noisecritique} identify failures
of independence and Gaussian assumptions under shared keys and ciphertext reuse.
ILA~\cite{ila} proves functional correctness through quantitative typing with
noise and wraparound bounds. Our deterministic downstream bounds specialize these concerns
to the retained circuit.

Our compiler makes second components functions of independent public masks,
reserved seeds and fixed maps, not queries or the gallery secret. The bounded
honest family then permits simultaneous Bernstein and union bounds over adaptive
queries and slots, without assuming independent convolution coefficients or
secret-coefficient independence after public observations. Composing this event with completed split
rounding and one-use current authority permits a smaller gallery modulus at an
explicit statistical cost to the ordinary Ring-LWE reduction. The contribution
is this composition, not retention, verification or concentration bounds alone.

\section{Discussion}
\label{sec:discussion}
\subsection{Security and Execution Scope}
The reductions assume faithful source and execution, exclusive nonrollback
owner custody, truthful history, and the declared entropy/XOF interfaces.
A deployed executor requires sound enrollment and confined computation/signing.
Local processes establish neither hardware attestation nor protection against
physical or microarchitectural observations of private variable-time arithmetic.

Lifetime budgets include failed and unused key families and reservations.
Worker reconstruction, owner restarts and additional controllers cannot reset
them. Public recomputation creates no fresh reservation. Concrete public-XOF
and private-sampler realizations require justification. Computational security
rests on the actual source-modulus Ring-LWE games and lifetime budgets,
not statistical tails.

\subsection{Performance and Ethics}
Permitted plaintext caching remains a faster cold-service alternative.
Historical-completion and blocked-expansion measurements retain separate scopes
(Appendices~\ref{app:editcommuting} and~\ref{app:earlier}).

Experiments use synthetic rows and local workloads, not personal records or
production services. Outputs are authorized only to an honest owner under the
stated model. Deployment outside it can expose sensitive information. These
checks do not establish production assurance.

\section{Conclusion}
The retention invariant and independent-recipient reduction preserve one-use
completion and exact current release. Second-component independence permits
simultaneous bounds for completed split joins at a smaller gallery modulus.
The implementation reduces end-to-end two-query cost despite slower preparation
and private completion, under the stated security and measurement scope.
\label{LastMainPage}

\raggedbottom
\bibliographystyle{IEEEtran}
\bibliography{references}

\appendices
\section{Exact Sampling and Public-XOF Simulation}
\label{app:sampling}
We match the exact sample distributions to Ring-LWE, then simulate complete public
byte streams with bounded runtime. Section~\ref{sec:security} states the
games, hybrid order, and bounds.

\subsection{CBD and Scaled Samples}
The noise distribution and invertible scaling map specify the oracle-to-ciphertext correspondence.
For an arbitrary positive integer $\eta$, a coefficient has
\[
 \Pr[e=j]=2^{-2\eta}\binom{2\eta}{\eta+j},\qquad -\eta\le j\le\eta.
\]
Indeed $\sum U_i-\sum V_i=\sum U_i+\sum(1-V_i)-\eta$ is a centered
$2\eta$-bit binomial. Independent disjoint bit blocks give the joint product distribution.
The secret is coefficientwise uniform ternary, including the possibility
that independently drawn role secrets have equal values.
Role namespaces are distinct. Secret values are not conditioned to differ.

Since $t$ is invertible modulo $Q$, multiplication of both ordinary sample
components by $t$ is a bijection on $R_Q^2$. It maps
$(a,-as+e)$ to $(ta,-(ta)s+te)$ and uniform pairs to uniform pairs.
It therefore gives the scaled-error sample reduction exactly.

\subsection{Complete Conditional Byte Streams}
Conditioning on accepted coefficients must preserve rejected bytes and every later XOF read.
Let $b_Q=\operatorname{bitlength}(Q)$ and
$p_{\rm acc}=Q/2^{b_Q}$. A width-$w$ candidate word, with $w\ge b_Q$,
contributes its low $b_Q$ bits.
Values at least $Q$ are rejected. For a prescribed accepted coefficient
$c\in[0,Q)$, sample a geometric rejection count $J$ with
$\Pr[J=j]=(1-p_{\rm acc})^jp_{\rm acc}$, then $j$ independent rejected low-bit values
uniform in $[Q,2^{b_Q})$. Include independent discarded high bits for
every word, including the accepted word, and an independent suffix after
the final accepted coefficient.

If the accepted coefficient is also uniform, the probability of a segment
with $j$ specified rejects and one specified acceptance is
\[
 (1-p_{\rm acc})^jp_{\rm acc}(2^{b_Q}-Q)^{-j}Q^{-1}
   =(2^{b_Q})^{-(j+1)}.
\]
Including high bits gives exactly the original uniform-word probability.
Repeat for all $N$ coefficients and append the suffix.
A direct realization draws candidate words, keeps their rejected words,
and replaces only the low bits of the first accepted word by the prescribed
$c$. This generates the same conditional stream without a second rejection
routine for the rejected interval. All later XOF reads see this complete
stream. Programming accepted words alone would omit observable randomness.
For direct-$P$ samples, the same construction uses $P$ in place of $Q$.

The domain includes the fixed tag, public role identity, and fresh seed.
Over $P_s$ fresh invocations and $h$ adversarial inputs, the union bound
for prequeries and seed collisions is Equation~\eqref{eq:xofbad}.
Gallery source expansions count even when only the expanded polynomial
is published. Republishing an identical asset does not allocate a new
programmed input.

\subsection{Finite Runtime and Sampling Cutoffs}
Bounded sampling introduces an abort loss without biasing successful values.
Because $Q>1$ is odd, $p_{\rm acc}>1/2$, and each accepted coefficient needs fewer
than two candidates in expectation. Capping every rejection stage at $\rho$
costs at most $2^{-\rho}$ per stage. The conservative
$C_{\rm samp}$ counts both uniform components in all-uniform samples, each
conditional parent stream, and secret draws. Apply this coupling once before
the recipient chain, with the same distribution of cutoff events in every hybrid.
For a finite-set rejection sampler, the abort probability depends on
acceptance probability, not on the eventual accepted value.
The capped interface can therefore draw that exact value and independently
abort, even at an unknown-secret challenge interface.

A fixed-work ternary-key sampler may scan $b_s$ two-bit candidates per coefficient,
using the first candidate below three and aborting if none exists.
For each $x\in\{-1,0,1\}$ its subprobability is
\[
 \Pr[\text{output }x]=(1-4^{-b_s})/3.
\]
This is exactly uniform ternary coupled to an independent exhaustion event.
The union bound over $R_s$ fresh role-secret draws is
$R_sN2^{-2b_s}$. For the concrete choice $b_s=96$, 16 draws, and
$N=2^{14}$, it is $2^{-174}$.
When this interface is used, account for its coupling within the sampler
term or as a separate term, once. OS entropy failure and private execution
are governed by their explicit interface assumptions.

\section{Key-Norm Cap Bounds}
\label{app:keycaps}
\subsection{Parameterized Ternary-Key Bounds}
Support and spectral tails certify wire profiles over a finite key lifetime.
Choose thresholds $K_s>2N/3$ for $\norm{s}_1$ and $U_s>0$ for
$\max_{\omega^N=-1}|s(\omega)|$, with $x_s=3U_s^2/(2N)>1$. Define
$J_s=\ceil{\sqrt{NK_sU_s^2}}$ and set
\begin{equation}
 (\Lambda_0^\star,\Lambda_1^\star,\Lambda_2^\star)
 =(1+K_s,\ 1+2K_s,\ 1+K_s+J_s).
 \label{eq:caps}
\end{equation}
The support and spectral failure budgets are
\begin{align}
 \delta_H&=\exp\!\left(-2(K_s-2N/3)^2/N\right),\notag\\
 \delta_{\rm sp}&=Nx_s\exp(1-x_s).
 \label{eq:tailbudgets}
\end{align}
These quantities depend on the key distribution and the thresholds, rather
than on the plaintext modulus or the source ciphertext modulus.

\begin{lemma}[Cap-failure bounds]
\label{lem:tails}
For fresh ternary roles with independent uniform coefficients, the pair,
mixed, and tensor cap-failure probabilities satisfy
\begin{equation}
 \delta_0\le\delta_H,\quad
 \delta_1\le2\delta_H,\quad
 \delta_2\le\delta_H+\delta_{\rm sp}.
 \label{eq:captails}
\end{equation}
The mixed-type union bound does not require independence between its two
role secrets, provided each has the stated marginal distribution.
\end{lemma}
\begin{proof}
For one ternary secret, $H_s=\norm{s}_1=\norm{s}_2^2$ is
$\operatorname{Bin}(N,2/3)$, so Hoeffding gives
$\Pr[H_s>K_s]\le\delta_H$. This proves pair and, by a union bound, mixed.
The Gaussian exponential argument in Appendix~\ref{app:tails} gives
\begin{equation}
 \Pr\!\left[\max_{\omega^N=-1}|s(\omega)|>U_s\right]
 \le Nx_s e^{1-x_s},\qquad x_s=\frac{3U_s^2}{2N}.
 \label{eq:spectral}
\end{equation}
On $H_s\le K_s$ and the spectral event's complement,
$\norm{s^2}_1\le\sqrt{NK_sU_s^2}\le J_s$ by the norm implication
proved in Appendix~\ref{app:tails}.
The two bad events have union probability at most $\delta_H+\delta_{\rm sp}$,
proving the tensor-type bound without assuming their independence.
\end{proof}

\paragraph{Selecting a statistical error budget}
For any target $\lambda>0$, sufficient conditions for
$\delta_\nu\le2^{-\lambda}$ in all three modes are
\begin{align}
 K_s&\ge\frac{2N}{3}+
       \sqrt{\frac{N(\lambda+1)\ln2}{2}},\notag\\
 x_s-1-\ln x_s&\ge\ln N+(\lambda+1)\ln2.
 \label{eq:statisticaltarget}
\end{align}
They make each of $\delta_H$ and $\delta_{\rm sp}$ at most
$2^{-(\lambda+1)}$. If a lifetime exposes at most $F_{\max}\ge1$
allowed fresh family/type pairs, replacing $\lambda$ on the right-hand sides by
$\lambda+\ceil{\log_2 F_{\max}}$ makes their total cap term at most
$2^{-\lambda}$. Here $\lambda$ controls statistical profile leakage,
and the Ring-LWE advantage remains a separate term.

\section{Ternary Spectral Tail}
\label{app:tails}
We bound the secret at negacyclic roots, then transfer that bound to the
coefficient norm of its negacyclic square.
For $j\ge1$, the coefficient inequality
$2\,3^{j-1}j!\le(2j)!$ holds at $j=1$ and is preserved inductively.
Comparison of even power-series coefficients yields
$(1+2\cosh z)/3\le\exp(z^2/3)$.

Write $s(X)=\sum_{i=0}^{N-1}s_iX^i$ and let $H_s=\norm{s}_1$ count its nonzero
coefficients. Use support threshold $K_s$, spectral threshold $U_s$ and norm
cap $J_s=\ceil{\sqrt{NK_sU_s^2}}$ for $s^2$, as in Appendix~\ref{app:keycaps}.
For $\omega^N=-1$ and $N$ even, $\sum_i\omega^{2i}=0$.
Thus the real and imaginary weight vectors are orthogonal with squared
norms $N/2$. For $Z=(\operatorname{Re}s(\omega),\operatorname{Im}s(\omega))$
and $h\in\mathbb R^2$, independence of $s_i$ gives
\[
 \mathbb E \exp(\langle h,Z\rangle)
 \le\exp(N\norm{h}_2^2/6).
\]
For a standard two-dimensional Gaussian $g$ and
$0<\theta<3/(2N)$, Tonelli and the Gaussian moment identity give
\begin{align*}
 \mathbb E_Z e^{\theta\norm{Z}_2^2}
 &=\mathbb E_g\mathbb E_Z
       e^{\sqrt{2\theta}\langle g,Z\rangle}\\
 &\le\mathbb E_g e^{N\theta\norm{g}_2^2/3}
   =(1-2N\theta/3)^{-1}.
\end{align*}
With $x_s=3U_s^2/(2N)>1$ and
$\theta=(1-1/x_s)/(2N/3)$, Markov yields
$x_se^{1-x_s}$ per root. A union bound over the $N$ roots gives
Equation~\eqref{eq:spectral}. Discrete Parseval gives
$\norm{s^2}_2\le\max_\omega|s(\omega)|\,\norm{s}_2$, using the exact integer
negacyclic square. Together with Cauchy--Schwarz, on $H_s\le K_s$ and
$\max_\omega|s(\omega)|\le U_s$, this yields
\begin{align*}
 \norm{s^2}_1&\le\sqrt N\,\norm{s^2}_2
 \le\sqrt N\,\max_\omega|s(\omega)|\,\norm{s}_2\\
 &\le\sqrt{NK_sU_s^2}\le J_s.
\end{align*}
This proves Lemma~\ref{lem:tails}'s norm implication without treating square
coefficients as independent.

\subsection{Numerical Certificate for the Experimental Thresholds}
For $N=16384$, $K_s=11820$, and $U_s=1100$, the coefficient-norm exponent
satisfies $2(K_s-2N/3)^2/N>141(347/500)$.
Also $x_s-1>158(347/500)$, $Nx_s<(15/16)2^{21}$, and
$\ln2<347/500$. The spectral tail is consequently below
$(15/16)2^{-137}$, while $\delta_H<2^{-141}=(1/16)2^{-137}$.
Equation~\eqref{eq:captails} then gives
$\delta_0<2^{-141}$, $\delta_1<2^{-140}$ and $\delta_2<2^{-137}$.
These rational comparisons instantiate the general theorem and
are not hypotheses of its proof.

\section{Terminal Conversion: Interpretation and Illustration}
\label{app:codec}
The terminal error budget combines scaled source phase and key-amplified rounding.
For a terminal with multipliers $\boldsymbol w$, write its coefficientwise
rounding errors as $\epsilon_i=r_p(C_i)-(p/q_c)C_i$.
Negacyclic multiplication satisfies
\[
 \norm{\epsilon_iw_i}_\infty
 \le\norm{\epsilon_i}_\infty\norm{w_i}_1
 \le(t/2)\norm{w_i}_1.
\]
Summing over components gives the rounding contribution $t\Lambda/2$.
The converted phase therefore has bound
$pB/q_c+t\Lambda/2$, rounded upward conservatively in
Equation~\eqref{eq:terminalbound}. For mixed and tensor terminals,
respectively, $\Lambda=1+\norm{s_g}_1+\norm{s_f}_1$ and
$\Lambda=1+\norm{s_g}_1+\norm{s_g^2}_1$.
The square is the integer negacyclic product. For ternary $s$,
$\norm{s}_1$ counts nonzero coefficients.

Ignoring the ceilings only to interpret the tradeoff, the strict guard is
equivalent to $p>t\Lambda/(1-2B/q_c)$. Smaller admitted key norms permit
smaller response moduli. Actual profile selection must also satisfy the
ceilings and $p\equiv q_c\bmod t$. The original keys, encrypted index, and
source modulus $q_c$ remain unchanged.

For an arithmetic illustration, take $q_c=1009$, $t=17$, and $B=100$.
These values are neither proposed security parameters nor benchmark data.
Both $p=57$ and $p=91$ have the same residue as $q_c$ modulo $t$:
\begin{center}
\begin{tabular}{rrrrrl}
\toprule
$p$ & $\Lambda$ & $\ceil{pB/q_c}$ & $\ceil{t\Lambda/2}$ & $B'$ & Guard \\
\midrule
57 & 2 & 6 & 17 & 23 & $23<28.5$ \\
57 & 3 & 6 & 26 & 32 & $32\not<28.5$ \\
91 & 3 & 10 & 26 & 36 & $36<45.5$ \\
\bottomrule
\end{tabular}
\end{center}
Thus the smaller modulus is certified at $\Lambda=2$, while $\Lambda=3$
requires a larger wire profile for this bound. Failure of the sufficient guard
does not demonstrate an incorrect plaintext. A certified fallback instead
retains exactness, using seven coefficient bits at $p=91$ rather than six
at $p=57$.

The same values also illustrate residue-preserving rounding. For $c=100$
and $p=57$, the scaled target is $(p/q_c)c\approx5.65$. Eligible integers
are congruent to $15$ modulo $17$, and the nearest is $r_p(c)=-2$.
Its error is about $7.65<17/2$. The stored value is $55$ modulo $57$,
whose residue modulo $17$ differs from that of $c$.
Lemma~\ref{lem:round} preserves the plaintext through the complete phase
and its carries, rather than through individual stored residues.
\section{Supported Blocked Expansion Graph}
\label{app:joingraph}
We unfold expansion blocks, track injected errors, and verify second-component independence.
The smaller-modulus join does not require a new generic expansion compiler.
The implemented blocked graph is obtained by unfolding adjacent levels of
the sequential trace and extraction circuit. A block spanning $b\in\{1,2\}$ levels
has one independent new recipient $s_v$. Its output maps on $f\in R_Q$ are
\[
 F_h(f)=\sum_u w_{h,u}\sigma_u(f),\qquad
 \sum_u\norm{w_{h,u}}_1\le 2^b,
\]
where $h$ indexes outputs, $u$ indexes automorphisms, $w_{h,u}$ is a signed
sum of monomials and $\sigma_u$ is a public negacyclic automorphism.
Unfolding transforms earlier shifts under later
automorphisms, preserving the original output order. The concrete compiler
checks the displayed coefficient mass on every assembled map.

For each required $\sigma_u$, switch directly from the block's input
recipient $s$ to $s_v$, with target $\sigma_u(s)$. Hoisting means reusing
one decomposition $C_1=\sum_{j<k_Q}\beta_Q^jd_j\bmod Q$ across branches,
with canonical coefficientwise digits $d_j$. Apply $\sigma_u$ to these digits. They reconstruct
$\sigma_u(C_1)$ modulo $Q$, and their coefficient magnitudes remain at most
$\beta_Q-1$. They need not be the canonical digits of the transformed
polynomial. No decomposition is distributed over a sum. A switched pair has
phase $\sigma_u(\Phi_s(C))+\epsilon_u$ with
$\norm{\epsilon_u}_\infty\le S_Q$. Public assembly gives
\[
 \Phi_{s_v}(\widehat C_h)
 =F_h(\Phi_s(C))+\sum_u w_{h,u}\epsilon_u\pmod Q.
\]
Here $\widehat C_h$ is the assembled output. If $V$ bounds previously injected
switching errors, its next bound $V'$ satisfies
$V'\le 2^b(V+S_Q)$. For blocks of width one or two, this is at most the
sequential recurrence $2^bV+2(2^b-1)S_Q$. Beginning at zero injected
switch error gives total propagated error at most $2(N-1)S_Q$.

This tracks only the injected errors. It does not claim that incomplete
sum-only algebraic traces, sparse restored phases or zero-first leaves have a fresh
phase bound. The complete unfolded coefficient operator equals the
sequential operator: it cancels uncontrolled projected coordinates and
multiplies the selected coordinate by $N$. The selected honest query
coefficient has magnitude at most $F_o$, giving the complete bound
$E_Q=NF_o+2(N-1)S_Q$ used in Theorem~\ref{thm:joincomposition}.

Second-component computation decomposes only the input second component and combines
transformed digits with hint second components. Public assembly does not read first components.
Induction across blocks thus establishes
Equation~\eqref{eq:joinnoninterference}. First-component linearity and the
complete extraction operator establish the retained affine identity.
Composite hint targets involve an earlier recipient only. Every independently
sampled column and zero parent is counted in the actual recipient's sample
budget, including fresh direct-$P$ gallery and final-hint samples. Changing
the block graph changes these budgets and its enrollment. It is not an
unaccounted reuse of the sequential sample interface.

\paragraph{Exact quotient residuals}
For an integer coefficient $c$ at $Q=Pa$ with odd $a\equiv1\pmod t$,
the rounding map has the equivalent
integer quotient form
\[
\begin{aligned}
 u&=\operatorname{center}_a((c\bmod a)t^{-1}\bmod a),\\
 r_{Q\to P}(c)&=(c-tu)/a.
\end{aligned}
\]
Here $\operatorname{center}_a$ selects the unique representative in
$[-(a-1)/2,(a-1)/2]$, and $t^{-1}$ is taken modulo $a$.
The numerator is divisible by $a$. The result is congruent to $c$ modulo
$t$, since $a\equiv1\pmod t$, and its residual is $-tu/a$ with magnitude
at most $t/2$. Thus $\rho_i$, $S_2$ and $M$ in
Lemma~\ref{lem:joinreservoir} are public integer computations on the actual
second component. Per-reservation moments may tighten an enrolled certificate,
but the measured instance uses a uniform worst-residual bound over the
whole honest reservoir. It does not assume uniform residuals or independent
convolution coefficients, nor infer a concrete Ring-LWE hardness level from $\delta_J$.

\paragraph{Experimental complete bound}
The measured construction profile uses $N=16384$, $d=512$, $t=1031$, $\eta=21$,
$H=16$, $C=10$ and $\ell_J=111$, with
\[
\begin{aligned}
 P&=1152921504606748673,\\
 a&=72057593002196993,\qquad Q=Pa.
\end{aligned}
\]
The uniform residual bound $\rho_i\le516$ gives $Z_*=822828$ and
$\delta_J<2^{-128}$. Source radix $2^{30}$ and final radix $2^{20}$ give complete-leaf bound
$E_J=1516649$ and joined-pair bound $B_J=277700936789622784<P/2$.
Final radix $2^{30}$ fails the bound and is excluded. These are statistical
correctness allocations, not computational security estimates. The 24-bit
terminal modulus selected by the measured P60 profile is $p=16767941$.
Its public-residual certificate is evaluated on the actual complete pair.

\section{Public Terminal Residual Certificates}
\label{app:residual}

An alternative certificate addresses the terminal rounding loss directly,
without changing the complete source-phase bound. It uses the actual public
ciphertext coefficients, rather than a tighter estimate of earlier hint errors.
The ingredients are residue-preserving modulus switching and an elementary
bound on sums of at most $K$ residual entries. The proposed extension concerns their typed composition
with historical evaluation and current one-use release.
High-rate constructions based on linear decryption and conversion to additive
encryption already address response compression~\cite{lineardecryption,additivecompression}.
This extension preserves the existing recipient graph and does not claim an
optimal communication rate or a new scheme-conversion primitive.

For a complete pair $c=(c_0,c_1)$, use canonical coefficient lifts and write
its bounded phase and integer carry as
\[
 c_0+c_1s=v+q_ch,\quad h\in R,\quad \norm v_\infty\le B<q_c/2.
\]
Let $s$ be an integer ternary polynomial with at most
$K\in\{0,\ldots,N\}$ nonzero coefficients.
For a registered odd
$1<p<q_c$ satisfying $p\equiv q_c\pmod t$, let $x=(x_0,x_1)$ with
$x_i=r_p(c_i)$ the nearest
\emph{unreduced} residue-preserving integer lift of
Equation~\eqref{eq:round}. Define integer-scaled rounding residuals and their
magnitude bounds coefficientwise
\[
 r_i=q_cx_i-pc_i,\qquad \rho_i=\ceil{|r_i|/q_c}.
\]
For a nonnegative vector $a$, $\TopSum_K(a)$ sums its $K$ largest entries,
with $\TopSum_0(a)=0$. It is distinct from the search ranking operator $\Top$.
Two public converted-phase bounds are
\begin{align*}
 E_p^*&=\ceil{\bigl(pB+\norm{r_0}_\infty+
                     \TopSum_K(|r_1|)\bigr)/q_c},\\
 E_p&=\ceil{pB/q_c}+\norm{\rho_0}_\infty+
                         \TopSum_K(\rho_1).
\end{align*}
\begin{proposition}[Public residual certificate]
\label{prop:publicresidual}
$E_p^*\le E_p\le E_p^*+K+1$. If $2E_p<p$, the pair $x\bmod p$ has centered
phase bounded by $E_p$ and the same plaintext as $c$.
\end{proposition}
\begin{proof}
The exact integer-ring identity is
\[
 x_0+x_1s-ph=(pv+r_0+r_1s)/q_c.
\]
Each coefficient of $r_1s$ is a signed sum of at most $K$ distinct entries
of $r_1$: negacyclic shifts permute their absolute values. Its magnitude is
therefore at most $\TopSum_K(|r_1|)$. Replacing the residual
magnitudes divided by $q_c$ by their ceilings adds less than $K+1$. The scaled
source ceiling adds less than one. Comparing the resulting integer ceilings
gives the displayed slack. The strict guard gives unique centering. Since
$x_i\equiv c_i\pmod t$ and $p\equiv q_c\pmod t$, subtraction of the same
integral carry $h$ preserves the plaintext residue.
\end{proof}

With $K=N$, this certificate holds for every ternary recipient and requires
no private support predicate. Moreover, $\rho_i\le\ceil{t/2}$, so a histogram
with $1+\ceil{t/2}$ bins computes it in $O(N+t)$ time. A registered fallback
satisfying
\[
 2\bigl(\ceil{pB/q_c}+(K+1)\ceil{t/2}\bigr)<p
\]
makes selection a total public function on canonical ciphertext inputs,
including uniformized hybrid inputs. Correct decoding additionally requires
the stated phase and support bounds. For several output groups, select one registered wire profile that
passes every group and sign its policy, group bounds, exact body extent and
unchanged historical issue prefix as a whole.

\section{Completion Across Truthful Edits}
\label{app:editcommuting}
The baseline admission algorithm retires an original when its issue head is
older than the current owner head. We analyze an optional variant in which
truthful edits commute with an in-flight immutable-base computation. Exact
private repair and current-digest validation have established predecessors in
verifiable databases and incremental computation~\cite{benabbas,integridb,hyperproofs}.
The question here is whether repair can preserve an already issued projected
original without relaxing its historical binding or one-use completion rule.

Let original $o$ bind query $q_o$, immutable computation $\Base_o$, generation
$\Generation_o$, and historical issue head $\Head_o$. Its signed prefix remains
unchanged. At a serialized release head $\Head_e$, require the following:
\begin{enumerate}
 \item The complete result matches the exact signed original and its
 immutable base, terminal type, wire profile, and current generation.
 \item The owner spends $o$ before whole-result private completion. No
 competing cache answer has released it, and no retired original is restored.
 \item The private query and complete truthful history from $D_0$ to $D_e$
 remain in exclusive owner custody.
 \item Correction, stable ranking, and release execute under the same
 release/edit lock at $\Head_e$.
\end{enumerate}
Generation equality and unchanged assets remain mandatory. Only issue-head
equality is replaced by live exact correction authority. Reconstruction,
history loss, and a changed base still revoke pending completion.

\begin{proposition}[Edit-commuting completion]
\label{prop:editcommuting}
Under the arithmetic, source, execution, codec, and owner-state assumptions of
Theorem~\ref{thm:correctness}, the optional transition above releases the exact
current-search result at its release head. It invokes private completion at
most once per original, including across repeated coordinates and reversions.
\end{proposition}
\begin{proof}
An admitted result is the prescribed encrypted evaluation for the base
database, independently of its issue head. Exact whole-result decoding gives
$v_0(q_o)$. Applying the base-to-current correction~\eqref{eq:correction} to the complete truthful
history $\History_e$ gives $v_0(q_o)+\Delta_e(q_o)=v_e(q_o)$ by
Lemma~\ref{lem:correction}. The lock fixes the same $e$ for repair and
release, and ranking with ties broken by row index yields the exact current top three.
Consumption precedes private arithmetic. Edits never recreate a spent object,
while reconstruction and failure retain monotone retirement. Induction on
transitions proves at-most-once completion.
\end{proof}

\paragraph{Privacy interface}
The edit-only refinement introduces no new encrypted parent, recipient target,
or terminal type.
Admission still authenticates the historical issue prefix before private
arithmetic. Under causally leakage-compatible schedules for the changed
transition system, idealization of private completion supplies the exact
current ideal answer at release, and the descending recipient hybrids are
unchanged. The existing finite loss bound consequently uses the actual fresh
sample and profile-exposure budgets. This statement retains the execution,
nonrollback, and logical observation premises of the main theorem.
When used, the head's optional $h_e$ component remains a fresh opaque label,
not a hash of the edits. A public hash chain of
raw private edits would permit enumeration of low-entropy edits and does not
instantiate this interface.

For the split join, apply the same refinement only with the source sampling assumptions and
terminal hypotheses of Theorem~\ref{thm:joincomposition}. The simultaneous
$\delta_J$ event already covers the declared family and is charged once
across the trace, rather than anew for an edit or a repeated public evaluation.

\section{Earlier Representation and Lifecycle Evidence}
\label{app:earlier}
Earlier measurements used two different constructions. Configuration A used
same-key switching and therefore needs an auxiliary assumption for its hint
columns. Its timings do not instantiate the independent-recipient theorem.
A qualified held workload observed post-preparation costs of 12.467 s for
an immutable encrypted base with client-side edit correction, 23.667 s for
server-side updates under authenticated execution and 35.127 s for the mutable
server construction with additional randomized tensor checks. These costs cover
the complete subsequent workload, not a single query. These architecture observations include
different transport and protocol placement and do not isolate one mechanism.
Permitted cache-only service cost 263 ms cold and 55.0 ms prepared.

An earlier independent-recipient pair-service study used 32,768 rows,
eleven queries, edits and paced links at 64 MiB/s for setup and 512 KiB/s
online. One calibration and three qualified held cohorts checked 198 top-three
outputs, including 60 current native completions. Capped terminals saved
8 KiB per reply and improved fresh-query medians by 12.93--18.16 ms for
one authenticated full-index response and 5.99--13.41 ms for retained objects
framed as an ordered group aggregate. Both require complete authenticated
coverage before private completion. Cold-cost ordering was
inconsistent and plaintext caching had the lowest full offline cost. These
are wire-profile results, not timings for the smaller-modulus join.

A separate six-configuration study uses blocked expansion without a private
key-cap test and a public residual codec. It compares completion of an in-flight
query across edits, retirement and reissuance after an edit, and permitted
plaintext caching.
With a controlled post-issue edit and shared 1.25\,MB/s downlink (decimal units),
three qualified held cohorts give median first answers of 519.962 ms for
completion across edits and 1055.238 ms for retirement and reissuance. Each
completes privately once, but reissuance requires two public evaluations rather
than one. The paired first-answer ratio is 1.973, but end-to-end workloads take
32.451 and 32.171 s, so no end-to-end improvement is established. In the unpaced no-edit
condition, cache answers supersede every HE result, which is retired without
private completion. This lifecycle study uses a different construction profile
from the main P60 comparison. Their timings are not composed.

\end{document}